\documentclass[aps,prl,superscriptaddress,twocolumn,longbibliography]{revtex4-1}
\usepackage[utf8]{inputenc}

\usepackage{tikz-cd}
\usepackage{color}
\usepackage{dsfont}
\usepackage{bm}
\usepackage{mathtools}
\usepackage{amsthm}
\usepackage{amsfonts,amssymb,amsmath}
\usepackage[hidelinks]{hyperref}
\usepackage{physics} \usepackage{multirow}
\usepackage{graphicx}

\usepackage[export]{adjustbox}
\usepackage{verbatim}

\usepackage{pdfpages} 
\usepackage{pgffor} 

\usepackage{soul} 

\newcommand{\dif}{\mathrm{d}}

\makeatletter
\AtBeginDocument{\let\LS@rot\@undefined}
\makeatother
\def\supplementfilename{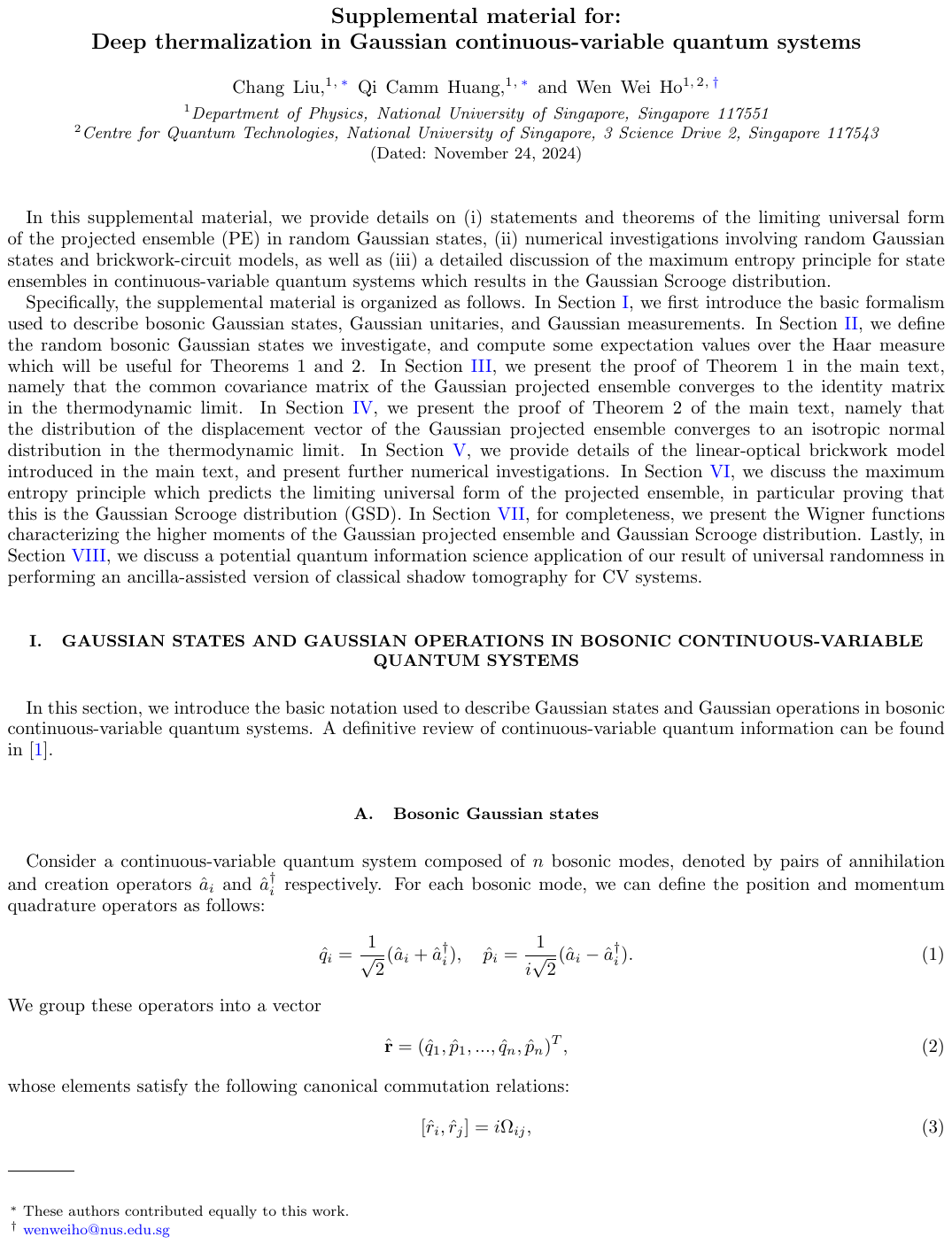}
\pdfximage{\supplementfilename}
\def\numbersupplementpages{\the\pdflastximagepages}

\begin{document}

\title{Deep thermalization in Gaussian continuous-variable  quantum systems} 
\author{Chang Liu}
\altaffiliation{These authors contributed equally to this work.}
\affiliation{Department of Physics, National University of Singapore, Singapore 117551}

\author{Qi Camm Huang}
\altaffiliation{These authors contributed equally to this work.}
\affiliation{Department of Physics, National University of Singapore, Singapore 117551}

\author{Wen Wei Ho}
\altaffiliation{\href{mailto:wenweiho@nus.edu.sg}{wenweiho@nus.edu.sg}}
\affiliation{Department of Physics, National University of Singapore, Singapore 117551}
\affiliation{Centre for Quantum Technologies, National University of Singapore, 3 Science Drive 2, Singapore 117543}

\begin{abstract} 
We uncover emergent universality arising in the equilibration dynamics of multimode continuous-variable systems. 
Specifically, we study the ensemble of pure states supported on a small subsystem of a few modes, generated by  Gaussian measurements on the remaining modes of a globally pure bosonic Gaussian state. 
We find that beginning from highly entangled, complex 
global states, such as random Gaussian states and product squeezed states coupled via a deep array of linear optical elements, the induced ensemble attains a universal form, independent of the choice of measurement basis: it is composed of unsqueezed coherent states whose displacements are distributed normally and isotropically, with variance depending on only the particle-number density of the system. 
We further show that the emergence of such a universal form is consistent with a generalized maximum entropy principle, which endows the limiting ensemble, which we call the ``Gaussian Scrooge distribution'', with a special quantum information-theoretic property of having minimal accessible information.
Our results represent a conceptual generalization of the recently introduced notion of ``deep thermalization'' in discrete-variable quantum many-body systems -- a novel form of equilibration going beyond thermalization of local observables -- to the realm of continuous-variable quantum systems.
Moreover, it demonstrates how quantum information-theoretic perspectives can unveil new physical phenomena and principles in quantum dynamics and statistical mechanics. 
\end{abstract}
\date{\today}
\maketitle

{\it Introduction.}---Identifying universal behavior exhibited by complex systems and simple, general principles behind their emergence is an important goal of physics. 
Quantum thermalization~\cite{
rigol2008thermalization, Nandkishore_review2015, RevModPhys.91.021001} is a prime example: under dynamics of generic isolated quantum many-body systems,  it is expected that local observables equilibrate to thermal values, governed only by global properties such as the conserved energy or charge.
Underpinning this is the relaxation of a local subsystem to a thermal Gibbs state 
due to the build-up of entanglement between the subsystem and its complement, whose appearance can be argued for appealing to the principle of maximal entropy in statistical physics~\cite{PhysRev.106.620}.

\begin{figure}[h]
\includegraphics[width=0.49\textwidth]{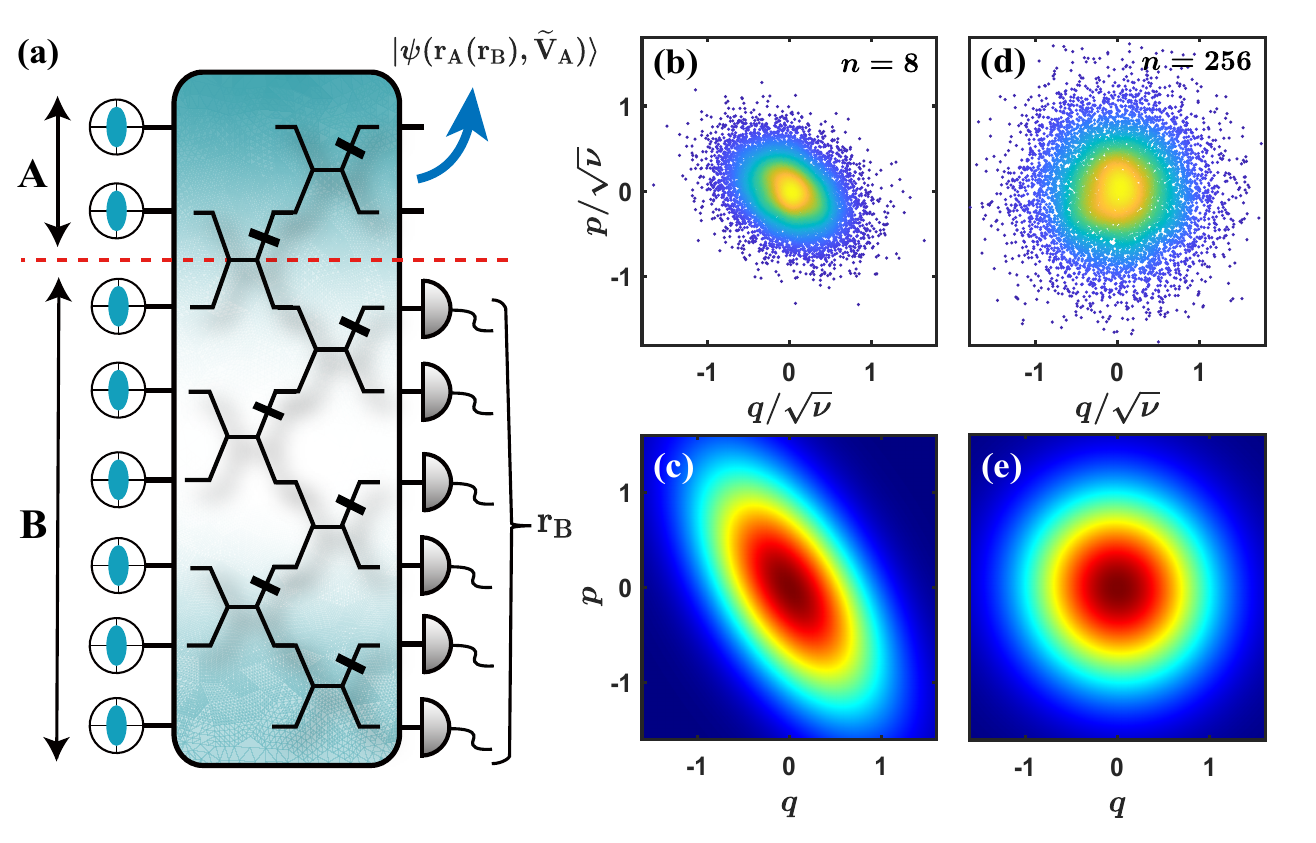}
\caption{(a) Deep thermalization in multimode CV quantum systems. 
Gaussian measurements on $n-k$ modes are performed on an $n$-mode global BGS (illustrated by squeezed light coupled via linear optical elements). 
This results in outcomes $\mathbf{r}_B$ and  projected pure BGS $|\psi(\mathbf{r}_A(\mathbf{r}_B),\widetilde{V}_A ) \rangle$ on the remaining $k$ modes, characterized by  displacement $\mathbf{r}_A$ and covariance matrix $\widetilde{V}_A$. 
 (b-c): At small $n$,  the distribution of $\mathbf{r}_A$ and $\widetilde{V}_A$ are both non-universal. (d-e): At large $n$, universality arises: $\mathbf{r}_A$ becomes distributed normally and isotropically with variance set only by particle-number density $\nu$, while $\widetilde{V}_A \to \mathbb{I}_A$. 
 Data was generated from the PE of a random BGS assuming coherent-state measurements.
}
\label{Fig:1}
\end{figure}

Recently, a new and stronger form of universality in the equilibration dynamics of strongly interacting, isolated quantum many-body systems was uncovered, dubbed {\it deep thermalization}~\cite{choi2023preparing,cotler2023emergent,PhysRevLett.128.060601,Ippoliti2023Dynamical, 
Ippoliti2022solvablemodelofdeep,Claeys2022emergentquantum,Shortriya2305, lucas2023generalized,Bhore2023deep,chan2024projected,varikuti2024unraveling,vairogs2024extracting,mark2024maximum, shaw2024}. 
This is the phenomenon of the  ensemble of pure conditional states of a local subsystem -- called the ``projected ensemble'' (PE)~\cite{choi2023preparing, cotler2023emergent,goldstein2016universal},  each of which is  tied to a measurement outcome of the complementary subsystem, acquiring  a {limiting universal} {\it distribution} 
over the Hilbert space 
which is maximally-entropic   according to a 
generalized  version of the second law of thermodynamics~(\cite{mark2024maximum}; and elaborated below).  
This can be understood as universality in the ``fine structure'' of the density matrix, namely its unraveling into constituent pure states according to measurements. 
For instance, in many-body systems of spins or fermions in the absence of conservation laws,   
 the  uniform or {Haar-distribution} emerges~\cite{
PhysRevLett.128.060601,Ippoliti2023Dynamical}; while   with conservation laws, it has been argued that the Haar-distribution distorted by the conserved charges, called the {Scrooge distribution}, emerges~\cite{cotler2023emergent, mark2024maximum}.  
In all of these past works, the finiteness of the local Hilbert space allows for a straightforward  derivation of the expected maximal-entropy state distributions. 
However, 
in systems of bosonic particles,    the unboundedness of the Hilbert space
poses a conceptual obstacle to immediately generalizing similar constructions (e.g., there is no normalizable uniform distribution to start from~\cite{Iosue2024_designs_CV}). 
It is thus an open question
what the form of deep thermalization achieved in 
{\it continuous-variable}~(CV) quantum systems is, and whether it follows similar maximal entropic principles. 

In this Letter, we consider the dynamics of multimode Gaussian CV quantum systems, and predict the limiting distribution attained by the PE under general conditions. 
Concretely, we focus on the collection of pure states supported on a few modes,  generated from Gaussian measurements of the complementary modes within a globally pure, entangled bosonic Gaussian state~(BGS)~(Fig.~\ref{Fig:1}a). 
We find that   
the PE constructed from a large class of highly entangled, complex 
global states~(elucidated below) tends to a universal form: it is composed of unsqueezed (and hence completely unentangled!) 
coherent states with displacements distributed normally and isotropically, with variance set only by the particle-number density. Remarkably, this happens independently of measurement basis. 
This is the manifestation of deep thermalization in CV systems.

At first sight, the phenomenology of deep thermalization between discrete-variable and CV systems seems surprisingly strikingly different: in the former, the post-measurement states -- 
being distortions of 
Haar random states -- are 
generically highly 
entangled; while in the latter, they are completely classical, with no quantum correlations whatsoever.  Despite these apparent differences, we further show that the limiting forms of the PE in these different physical systems are in fact governed by the common principle of having 
  quantum information-theoretic properties of possessing minimal {\it accessible information}~\cite{Jozsa1994lowerbound,Nielsen_Chuang_2010, mark2024maximum}. Thus, our work illustrates that the quantum-information rooted entropic principles underlying deep thermalization are powerful and general, and can be used to predict the emergent late-time universal behavior across distinct physical systems.

\textit{Projected ensemble in CV systems.}---We first recap the object of interest: the projected ensemble~(PE),  within the context of extended spin systems. Consider a global pure state $|\Psi\rangle$ on a bipartite system $AB$, and measurements  of $B$ (the `bath') in the local computational basis, which yield bit-strings $\mathbf{z}_B$ 
and corresponding pure projected states 
$|\psi(\mathbf{z}_B)\rangle$\,$=$\,$\hat I_A\otimes \langle\mathbf{z}_B|\Psi\rangle/\sqrt{p(\mathbf{z}_B)}$ on $A$, occurring with Born probability $p(\mathbf{z}_B)$\,$=$\,$\langle\Psi|({\hat I}_A\otimes |\mathbf{z}_B\rangle\langle\mathbf{z}_B|)|\Psi\rangle$. The PE is the set of all probabilities and associated conditional states 
$\mathcal{E}:=\{p(\mathbf{z}_B),|\psi(\mathbf{z}_B)\rangle\} $, and describes a {\it distribution} of pure states over the Hilbert space of $A$.
As mentioned, it has been found that  the PE attains a universal, `maximally-entropic' form constructed from  highly entangled, complex  
global states $|\Psi\rangle$, such as those arising in the dynamics of  quantum chaotic many-body systems~\cite{choi2023preparing,cotler2023emergent,PhysRevLett.128.060601,Ippoliti2023Dynamical, 
Ippoliti2022solvablemodelofdeep,Claeys2022emergentquantum,Shortriya2305, lucas2023generalized,Bhore2023deep,chan2024projected,varikuti2024unraveling,vairogs2024extracting,mark2024maximum, shaw2024};
this is a novel form of equilibration  called `deep thermalization' which goes beyond regular thermalization, as it constrains the late-time behavior of the full distribution of a local system (unraveled by measurement outcomes of the bath) and not just the expected values of its observables, which constitute the average behavior.

 In this work, we study the PE arising in  CV systems of $n$ bosonic modes.
We abide by a convention  in which the quadrature operators (i.e.,~position and momentum operators) are grouped as a vector $\hat{\mathbf{r}}$\,$=$\,$
(\hat{q}_1$\,$,$\,$\hat{p}_1$\,$,$\,$ \cdots$\,$,$\,$\hat{q}_n$\,$,$\,$\hat{p}_n)^T$ obeying the canonical commutation relations  $[\hat r_i$\,$,$\,$\hat r_j]$\,$=$\,$i\Omega_{ij}$, with
$\Omega$\,$=$\,${\bigoplus_{i=1}^n  \begin{pmatrix}
    0 & 1 \\
    -1 & 0
\end{pmatrix}
}
$
being the $n$-mode symplectic form~(see the Supplemental Material~(SM)~\cite{supp} for details).
We adopt a standard phase space representation of quantum states $\hat\rho$ via the Wigner function of the state
$W(\mathbf{x})$ where $ \mathbf{x}$\,$=$\,$ (q_1,p_1,...,q_n,p_n)^T \in \mathbb{R}^{2n}$,  
%
which is a real quasi-probability distribution~\cite{supp}. 
We henceforth focus on bosonic Gaussian states~(BGS), whose Wigner functions are multivariate Gaussian functions fully characterized by their first two statistical moments~\cite{weedbrook2012gaussian}: the displacement vector $\mathbf{r}$\,$:=$\,$\text{Tr}({\hat\rho}\hat{\mathbf{r}})$\,$\in$\,$\mathbb{R}^{2n}$, which we will always set to $\mathbf{0}$ without loss of generality, and the $2n$\,$\times$\,$2n$ positive covariance matrix $V_{ij}$\,$:=$\,$\text{Tr}\left(\hat\rho\{\mathbf{\hat{r}}_i-\mathbf{r}_i,\mathbf{\hat{r}}_j-\mathbf{r}_j\}\right)$, where $\{\cdot$\,$,$\,$\cdot\}$ is the anti-commutator.

We construct the PE on a subsystem $A$ of $k$ modes as the set of pure states arising from continuous positive operator-valued measures (POVM) on the complementary subsystem $B$ of $n$\,$-$\,$k$ modes. We will consider specifically Gaussian measurements parameterized by covariance matrix $\sigma_B$, defined by rank-1 projectors $\{\hat\Pi(\mathbf{r}_B,\sigma_B)\propto|\phi(\mathbf{r}_B,\sigma_B)\rangle\langle\phi(\mathbf{r}_B,\sigma_B)|\}_{\mathbf{r}_B}$, where $|\phi(\mathbf{r}_B,\sigma_B)\rangle$ is a BGS on $B$ with displacement $\mathbf{r}_B$ and covariance $\sigma_B$. Note that $\int \dif \mathbf{r}_B \hat\Pi(\mathbf{r}_B,\sigma_B)$\,$=$\,$\hat I_B$. 
A measurement yields an outcome $\mathbf{r}_B$ and an associated projected state $|\psi(\mathbf{r}_A(\mathbf{r}_B), \widetilde{V}_A)\rangle$ on $A$, which is also a BGS, with probability density $p(\mathbf{r}_B) $\,$\propto$\,$e^{-\mathbf{r}_B^T(V_B+\sigma_B)^{-1}\mathbf{r}_B}$. Here, the displacement and covariance of a projected state is~\cite{weedbrook2012gaussian}
\begin{align}
    & \mathbf{r}_A(\mathbf{r}_B)  =   V_{AB}(V_B+\sigma_B)^{-1}\mathbf{r}_B,\label{eqn:ram}
     \\ 
    & \widetilde{V}_{A} = V_A-V_{AB}(V_B+\sigma_B)^{-1}V_{AB}^T, \label{eqn:Vam}
\end{align}
where we have used the block-matrix decomposition of the covariance matrix of the global state  
\begin{equation}
    V = \left(
    \begin{array}{cc}
         V_A &V_{AB}  \\
         V_{AB}^T &V_B
    \end{array}
    \right)
\end{equation}
into correlations $V_{A(B)}$ within subsystems $A(B)$ and the correlations $V_{AB}$ in-between.  
Since $\mathbf{r}_A(\mathbf{r}_B)$ is a linear transformation of $\mathbf{r}_B$, the distribution of displacements $p(\mathbf{r}_A)$ follows a multivariate normal distribution 
$\mathbf{r}_{A}$\,$\sim$\,$\mathcal{N}(\mathbf{0}$\,$,$\,$\Sigma_A)$ with  covariance $\Sigma_A$\,$=$\,$\frac{1}{2}V_{AB}(V_B+\sigma_B)^{-1}V_{AB}^T$; while  $\widetilde{V}_A$ is  independent of measurement outcome. 
We thus see that the PE in a Gaussian CV system with Gaussian measurements can be compactly expressed as
\begin{equation}
    \mathcal{E}_G =\{p(\mathbf{r}_{A}),|\psi(\mathbf{r}_{A},\widetilde{V}_{A})\rangle\},
\end{equation}
which only requires two pieces of information: (i)~the distribution of displacements $\mathbf{r}_A$, captured by covariance matrix $\Sigma_A$; and (ii)~the (common) quantum correlations of quadratures of a projected state, captured by covariance matrix $\widetilde{V}_A$.  In principle, these depend on the measurement basis 
$\sigma_B$.

Our central claim is that for a large class of highly entangled, complex 
global BGS,  the PE  on  a fixed number of $k$ modes    acquires a remarkably simple universal form in the thermodynamic limit (TDL) $n \to \infty$, independent of measurement basis $\sigma_B$:  the projected states are all unsqueezed coherent states, whose displacements $\mathbf{r}_A$ are distributed normally and isotropically with variance set only by the particle-number density  of the system $\nu := \langle\hat{N}\rangle/n$ 
where $\hat{N}$
is the number operator, see Fig~\ref{Fig:1}(b-e).
Concretely, we claim 
\begin{equation}
   \mathbf{r}_{A} \stackrel{d}{\to} \mathcal{N}_{\nu} := \mathcal{N}(\mathbf{0},\nu \mathbb{I}_A),
     \quad \widetilde{V}_A\rightarrow\mathbb{I}_A,
     \label{eqn:central_result}
\end{equation}
where $\mathbb{I}_A$ is the identity matrix of size $2k$. 
In what follows, we will support our claim of the emergence of such universality through a combination of rigorous analytical statements and extensive numerical investigations, considering both random BGS and a physical model of squeezed light passed through a (fixed) array of beam-splitters and phase-shifters, as well as across different measurement bases. 

{\textit{PE from random BGS.--}
Consider first a random pure BGS on $n$ modes.
Precisely, as we can   write 
a pure BGS's covariance matrix as
$V$\,$=$\,$O(\bigoplus_{i=1}^n Z_i) O^T$ where  
$Z_i$\,$=$\,$\text{diag}(e^{2s_i}, e^{-2s_i})$ ($s_i$\,$\in$\,$\mathbb{R}$) and 
$O$ belongs to the real ortho-symplectic group $\text{Sp}(2n,\mathbb{R})$\,$\cap$\,$\text{O}(2n,\mathbb{R})$ which is isomorphic to the complex unitary group $U(n)$~\cite{weedbrook2012gaussian}, 
the random BGS we consider will be defined as those where $s_i$\,$=$\,$s$ are fixed and $O$ is drawn uniformly from the Haar measure (on the ortho-symplectic or unitary group).
Physically, $Z_i$ represents the covariance matrix of a one-mode squeezed vacuum with squeezing parameter $s_i$, while $O$ represents a passive (particle-number conserving) Gaussian unitary. Our choice of random states thus corresponds to uniformly-squeezed product states evolved via a Haar random particle-number conserving unitary, such that the particle-number density $\nu$\,$=$\,$(\cosh(2s)$\,$-$\,$1)/2$ is fixed and well-defined in the TDL. We note that entanglement properties ~(e.g.,~Page curves)
of such random BGS have recently been studied~\cite{Serafini_2007,
Fukuda2019typical,Iosue2023pagecurvestypical}.

We construct the $k$-mode PE on $A$ assuming coherent-state measurements $\sigma_B$\,$=$\,$\mathbb{I}_B$ on $B$. Our first result pertains to the limiting form of the covariance matrix of the Wigner function of each projected state: \\
{\bf Theorem 1.} {\it Consider $\widetilde{V}_A$,  the (common) covariance matrix  of the projected states on $k$-modes, generated from coherent-state measurements on the complement of a random $n$-mode BGS. For any $\epsilon$\,$>$\,$0$,
\begin{align}
\mathbb{P}( \| \widetilde{V}_A - \mathbb{I}_A \|_1 \geq \epsilon) \leq C(1+\epsilon/(2k))/\epsilon^2 n,
\label{eqn:Thm1}
\end{align}
 where $C$ is a constant depending on $k,s$ but not $n$. Here $\|\cdot\|_1$ is the trace norm.  
}\\
Theorem 1 establishes that with unit probability in the TDL, the projected states of a random BGS are all unsqueezed coherent states on $A$. The proof proceeds using a Makov-like inequality from probability theory to bound the deviation of $\tilde{V}_A$ from $\mathbb{I}_A$ by its expected value, and Weingarten calculus (integration over the unitary group) to estimate the latter; details are provided  in~\cite{supp}.
Further, using similar techniques as well as  concentration of measure statements on unitary groups (c.f.~Levy's lemma~\cite{anderson2010introduction}), we can characterize the convergence of the distribution of displacements of the projected states: \\ 
{\bf Theorem 2.} {\it Consider $\mathcal{N}_A$\,$=$\,$\mathcal{N}(0,\Sigma_A)$, the distribution of displacements $\mathbf{r}_A$ of the projected states on $k$-modes, generated from coherent-state measurements on the complement of a random $n$-mode BGS. For any $\epsilon > 0$,
\begin{align}
    \mathbb{P}( D_\mathrm{KL}(\mathcal{N}_A || \mathcal{N}_\nu) \geq \epsilon ) \leq 
    C \frac{1+ \frac{\nu \epsilon'}{2k}}{\nu^2 \epsilon'^2 n }
    + 
    D e^{-n \nu^2 \epsilon'^2 F},
\end{align}
where $C,D,F$ are constants depending on $k,s$ only,  and $\epsilon' = \frac{1}{2} \left( \sqrt{\epsilon/k} \sqrt{4 + \epsilon/k} - \epsilon/k \right)$.
$D_\mathrm{KL}$ is the Kullback-Leibler divergence of $\mathcal{N}_A$ with respect to $\mathcal{N}_{\nu}$~\cite{supp}.
}\\
Theorem 2 expresses that with unit probability in the TDL, the distribution of displacements $\mathcal{N}_A$ is statistically indistinguishable from an isotropic normal distribution $\mathcal{N}_\nu$ with variance $\nu$. 
 Together, Theorems 1 and 2 constitute a concrete realization of our claim of universality:  a {\it typical} random BGS is deep thermalized,  i.e.,~the PE constructed from it invariably has the limiting form  Eq.~\eqref{eqn:central_result}.

\begin{figure}[t]
\centering
\includegraphics[width=0.49\textwidth]{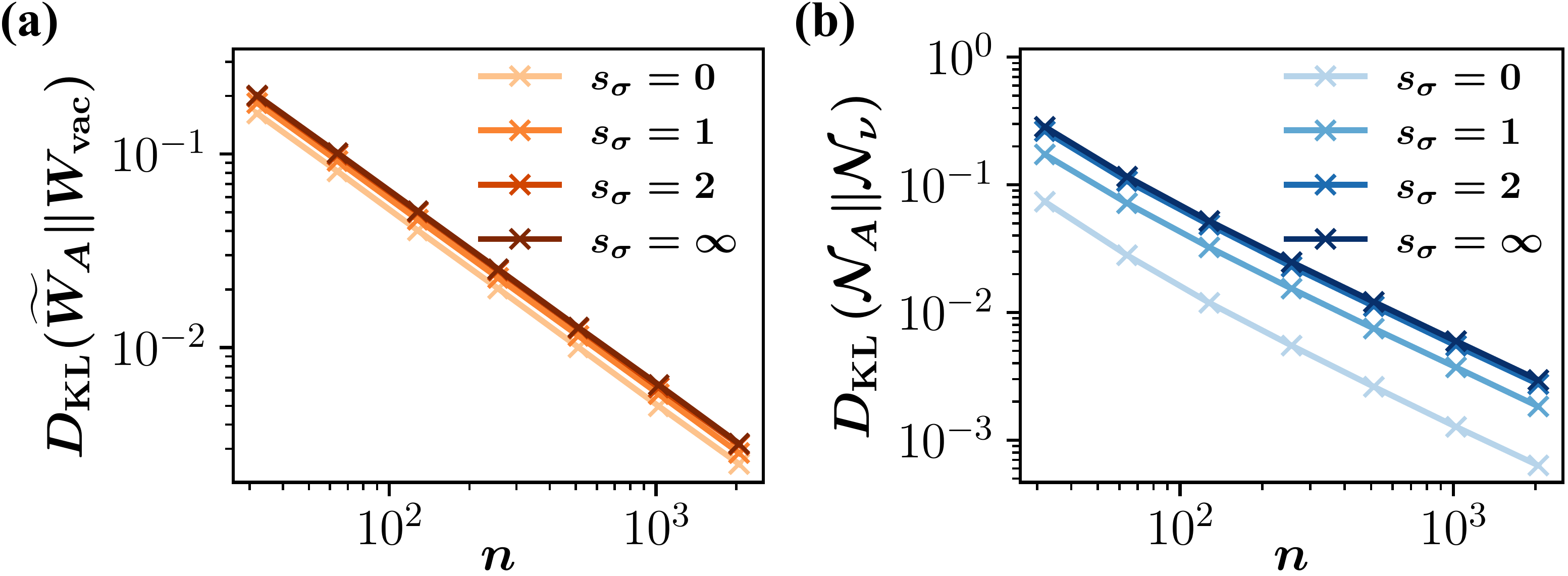}
\caption{Average KL-divergences of (a) the Wigner functions of a $3$-mode projected state from a coherent state of equal displacement, and (b) distribution of displacements  $\mathcal{N}_A$ from the expected distribution $\mathcal{N}_{\nu}$, for various product squeezed-state measurement bases parameterized by $s_\sigma$. The global state is a random $n$-mode BGS with uniform squeezing $s=1/2$. Note $s_\sigma$\,$=$\,$0$ corresponds to coherent-state measurements, $s_\sigma$\,$=$\,$\pm\infty$   to $\hat{x}/\hat{p}$ quadrature or homodyne measurements, and general $s_\sigma$ to general squeezed-state measurements. Both KL-divergences decay as $n^{-1}$.
}
\label{Fig:2}
\end{figure}
 However, we note that our rigorous results only pertain to coherent-state measurements $\sigma_B$\,$=$\,$\mathbb{I}_B$. An immediate question is the dependence of the emergent universality on general measurement bases. 
 To this end, we numerically simulate measurements in various  uniform product squeezed-state bases $\sigma_B$\,$=$\,$\bigoplus_{i=1}^{n-k}\text{diag}(e^{2s_\sigma},e^{-2s_\sigma})$ with  squeezing parameter $s_\sigma$.  In Fig.~\ref{Fig:2}(a) we plot the average KL-divergence of the Wigner functions of a projected state with a coherent state of identical displacement, 
  which captures the two states' statistical distinguishability; equivalently it probes the closeness of $\widetilde{V}_A$ to $\mathbb{I}_A$. In Fig.~\ref{Fig:2}(b) we plot the average KL-divergence of the distribution $\mathcal{N}_A$ of displacements $\mathbf{r}_A$ from the expected isotropic distribution $\mathcal{N}_\nu$. We find both distances always decay to zero in the TDL as $n^{-1}$, remarkably insensitively to the measurement basis.

\textit{PE from linear-optical circuit evolution.}---Next, we study  global states realized under more physical scenarios: namely, single-mode squeezed states coupled through an array of local passive Gaussian unitaries, (i.e., beam-splitters and phase-shifters).
This is reminiscent of the set-up in Gaussian Boson sampling (though a crucial difference is in the measurement bases)~\cite{Lund2014bosonsampling,Hamilton2017bosonsampling,kruse2019bosonsampling,grier2022complexity,Abhinav2022}.
Concretely, consider a linear array of $n$  bosonic modes each prepared in a squeezed vacuum with squeezing parameter $s$, and 
couple them by a $t$-layer (interpreted as `time') brickwork circuit where one layer consists of identical beam-splitter+phase shift operations on odd pairs of modes, followed by the same operations on even pairs of modes~(see Fig.~\ref{Fig:1}a). We then construct the PE on $k$-modes, which depends on time~(see SM for details~\cite{supp}).
In Fig.~\ref{Fig:3}(a-b) we plot the dynamical behavior of the KL-divergences of the Wigner function of the projected state and distribution of displacements from their expected limiting forms,
for a representative choice of fixed local beam-splitter and phase-shift, chosen uniformly across the circuit in space and time. Despite the evolved global state being non-random, we find again that both divergences vanish in the TDL and large circuit-depth limit, taken in that order.
In \cite{supp} we provide yet more numerics detailing the insensitivity of this outcome to other choices of system parameters and measurement bases. 
Together with our analytic and numerical results from the class of random BGS, these investigations strongly support the {\it universality} of the limiting form of the PE Eq.~\eqref{eqn:central_result}.

\begin{figure}[t]
\centering
\includegraphics[width=0.48\textwidth]{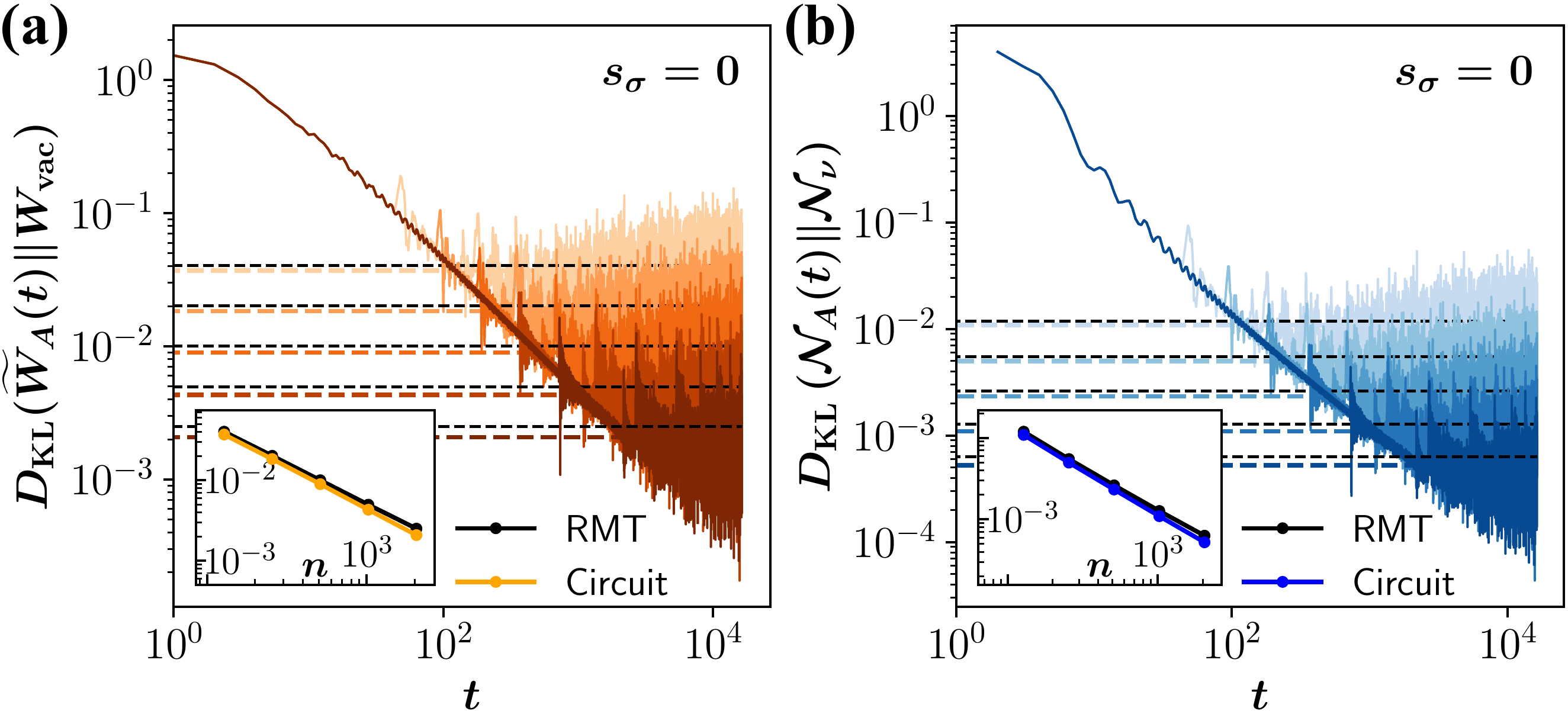}
\caption{KL-divergences over time of (a) Wigner functions of a $3$-mode projected state from a coherent state of equal displacement, and (b) distribution of displacements $\mathcal{N}_A$ from the expected distribution $\mathcal{N}_\nu$.
The projected state is generated from a $t$-layer brickwork interferometric circuit acting on an $n$-mode product squeezed state with uniform squeezing  $s=1/2$, for $n=128, 256, 512, 1024, 2048$ (light-dark;~`Circuit'). 
Both KL-divergences decay as $t^{-1}$ till they saturate to values (brown/blue-dashed) in very good agreement with those of  the corresponding projected ensemble
constructed from an $n$-mode random BGS (black-dashed;~`RMT'). Inset shows this agreement and the  saturation scaling  as $\sim n^{-1}$. 
}
\label{Fig:3}
\end{figure}

\textit{Maximum entropy principle and Gaussian Scrooge distribution}.---We now expound on the physical principles underlying the emergence of the universal ensemble. We first note that the PE can be understood as a particular (physically-motivated) 
unraveling of the reduced density matrix (RDM) $\hat\rho_A$, specified  by covariance $V_A$, into a collection of {\it pure} BGS with displacements distributed as $\mathcal{N}(\mathbf{0},\Sigma_A)$ and common covariance $\widetilde{V}_A$, such that $V_A$\,$=$\,$2\Sigma_A$\,$+$\,$\widetilde{V}_A$. 
In the TDL, one can easily argue for the limiting form of $\hat\rho_A$ using the standard principle of entropy-maximization in statistical physics~(subject to conservation laws; here, only particle number $\nu$):
it should be a thermal Gibbs state $\hat{\rho}_\text{th}$\,$\propto$\,$e^{-\beta \hat{N}_A}$ where $\nu^{-1}$\,$=$\,$(e^\beta-1)$, which has covariance $V_\text{th}$\,$=$\,$(2\nu+1)\mathbb{I}_A$. 
The entropy used in this maximization is the von Neuman entropy of $\hat\rho_A$, which is related to the minimization of the free energy of the local subsystem~\cite{Preskill229_Notes}. 

However, this max-entropy principle fixes only the limiting form of $V_A$, but not its constituents $\Sigma_A$\,$,$\,$\widetilde{V}_A$. 
To determine these, we propose a generalized maximum entropy principle that the limiting PE should obey.
%
First we argue that 
the entropy of a {\it state-ensemble} $\mathcal{E}$ is contained not only in the  distribution of constituent states, but also in the information one can extract via quantum measurements on them. To that end we define the ensemble entropy 
\begin{equation}
    \mathcal{S}(\mathcal{E}) := -\sup_{\mathcal{M} \in \text{POVM} }D_\mathrm{KL}( p_{\mathcal{E},\mathcal{M}} \|p_\mathcal{E}\otimes p_{\mathcal{M}})
    \label{eqn:S}
\end{equation}
where $\mathcal{M}$ runs over all POVMs, $p_{\mathcal{E},\mathcal{M}}$ is the joint distribution of states of $\mathcal{E}$ and measurement outcomes of $\mathcal{M}$ within them, and  $p_{\mathcal{E}},  p_{\mathcal{M}}$ the marginals. 
$-S(\mathcal{E})$ is in fact interpretable as the {\it accessible information} $I(\mathcal{E})$ of the ensemble~\cite{Nielsen_Chuang_2010}, which is the maximum amount of classical information extractable from measurements when classical data is encoded with quantum states of $\mathcal{E}$. 
In our context of the PE, this encoding arises due to the entanglement  between $A$ and $B$ within the global state (e.g., coming from dynamics). Now, if such entanglement is complex, then  information should be hidden well, and thus we can expect the limiting ensemble is one that minimizes $I(\mathcal{E})$ (or equivalently maximizes $S(\mathcal{E})$) across all possible ensembles $\mathcal{E}$, subject to the constraint that it has mean $\hat\rho_\text{th}$~(set by regular thermalization).


In Theorem 3 in the SM~\cite{supp}, utilizing recent seminal works by Holevo~\cite{holevoMaximizer, holevo2021accessible}, we show that the Gaussian state-ensemble 
satisfying this principle is 
\begin{equation}
\mathcal{E}_{\text{GSD}}= \left\{\dif \mathbf{r}_A
p(\mathbf{r}_A),     {\sqrt{\hat{\rho}_\mathrm{th}}|\mathbf{r}_A \rangle}/\sqrt{\langle  \mathbf{r}_A  |\hat{\rho}_\mathrm{th}|  \mathbf{r}_A  \rangle}\right\},
\label{eqn:GSE}
\end{equation}
where $p(\mathbf{r}_A)$\,$=$\,$\langle  \mathbf{r}_A  |\hat{\rho}_\mathrm{th}|  \mathbf{r}_A  \rangle/(2\pi)^k$  and $|\mathbf{r}_A\rangle$
is a (unsqueezed) coherent state on $A$ with displacement $\mathbf{r}_A$\,$\in$\,$\mathbb{R}^{2k}$, which we term the `Gaussian Scrooge distribution'~(GSD). 
We see this ensemble is composed of BGS with displacements distributed as $\mathcal{N}(\mathbf{0}$\,$,$\,$\nu\mathbb{I}_A)$ and covariance $\widetilde{V}_A$\,$=$\,$\mathbb{I}_A$ --- precisely our claim of deep thermalization Eq.~\eqref{eqn:central_result}!
%
 We note that interestingly,
 the emergence of the Scrooge distribution in spin systems~\cite{Jozsa1994lowerbound,goldstein2006distribution,reimann2008typicality,goldstein2016universal} can also be argued for using this generalized maximum entropy principle, so that it has a similar quantum information theoretic property of having minimal accessible information. Further, the forms of the GSD and Scrooge ensemble bear striking similarity:  they are both `$\hat\rho$-distorted' distributions, though of  differing underlying ensembles, namely uniform coherent states and  Haar random states respectively~\cite{supp}.

\textit{Discussion and outlook}.---Our work has uncovered a novel form of universality in the equilibration dynamics of Gaussian multimode CV quantum systems: the emergence of a maximally-entropic, minimally-information-yielding ensemble of local post-measurement Gaussian states called the Gaussian Scrooge distribution. This is the conceptual extension of the phenomenon of deep thermalization, originally formulated in spin or fermionic systems, for Gaussian CV systems. 
Intriguingly, the same fundamental quantum information-theoretic principle of maximization of entropy as the one discussed in this work has been found to underpin the emergent universality in spin systems too~\cite{mark2024maximum}, highlighting the generality and power of the principle across distinct physical systems. Moving forward, it would be very interesting to understand what this principle predicts for the universal form of the PE in other scenarios, such as if the particle-number density $\nu$ grows with system size $n$, or if we relax the assumption of Gaussianity (e.g., Fock or Gaussian states with Fock-state measurements). An analysis incorporating the effects of noise and loss, will also be important for an experimental verification of the phenomenon of Gaussian deep thermalization. 
It may also be possible to harness the universal randomness generated for quantum information science applications: for example,  the random Gaussian states generated on a target system upon measuring additionally introduced degrees of freedom, may be employed in an ancilla-assisted version of CV classical shadow tomography protocol based on random Gaussian unitaries and Gaussian measurements~\cite{becker2024classical}, extending similar ideas that have been recently proposed in spin systems~\cite{mcginley2023shadow,tran2023measuring}.




\begin{acknowledgments}
{\it Acknowledgments.}~We thank Soonwon Choi, Max McGinley, Hannes Pichler,  and Di Zhu for interesting discussions. W.~W.~H. is supported by the Singapore NRF Fellowship, NRF-NRFF15-2023-0008, and the CQT Bridging Fund. 
\end{acknowledgments}

 \bibliography{biblio}

 

\section*{Supplementary material}

Chang Liu,[1, *] Qi Camm Huang,[1, *] and Wen Wei Ho[1, 2, †]

[1]*Department of Physics, National University of Singapore, Singapore 117542*
[2]*Centre for Quantum Technologies, National University of Singapore, 3 Science Drive 2, Singapore 117543*
(Dated: May 8, 2024)

In this supplemental material, we provide details on (i) statements and theorems of the limiting universal form of the projected ensemble (PE) in random Gaussian states, (ii) numerical investigations involving random Gaussian states and brickwork-circuit models, as well as (iii) a detailed discussion of the maximum entropy principle for state ensembles in continuous-variable quantum systems which results in the Gaussian Scrooge distribution.

Specifically, the supplemental material is organized as follows. In Section I, we first introduce the basic notation used to describe bosonic Gaussian states, Gaussian unitaries, and Gaussian measurements. In Section II, we define the random bosonic Gaussian states we investigate, and compute some expectation values over the Haar measure which will be useful for Theorems 1 and 2. In Section III, we present the proof of Theorem 1 in the main text, namely that the common covariance matrix of the Gaussian projected ensemble converges to the identity matrix in the thermodynamic limit. In Section IV, we present the proof of Theorem 2 of the main text, namely that the distribution of the displacement vector of the Gaussian projected ensemble converges to an isotropic normal distribution in the thermodynamic limit. In Section V, we provide details of the linear-optical brickwork model introduced in the main text, and present further numerical investigations. In Section VI, we discuss the maximum entropy principle which predicts the limiting universal form of the projected ensemble, in particular proving that this is the Gaussian Scrooge distribution (GSD). Lastly, in Section VII, for completeness, we present the Wigner functions characterizing the higher moments of the Gaussian projected ensemble and Gaussian Scrooge distribution.

## I. GAUSSIAN STATES AND GAUSSIAN OPERATIONS IN BOSONIC CONTINUOUS-VARIABLE QUANTUM SYSTEMS

In this section, we introduce the basic notation used to describe Gaussian states and Gaussian operations in bosonic continuous-variable quantum systems. A definitive review of continuous-variable quantum information can be found in [1].

### A. Bosonic Gaussian states

Consider a continuous-variable quantum system composed of $n$ bosonic modes, denoted by pairs of annihilation and creation operators $\hat{a}_i$ and $\hat{a}_i^\dagger$ respectively. For each bosonic mode, we can define the position and momentum quadrature operators as follows:

$$\hat{q}_i = \frac{1}{\sqrt{2}}(\hat{a}_i + \hat{a}_i^\dagger), \quad \hat{p}_i = \frac{1}{i\sqrt{2}}(\hat{a}_i - \hat{a}_i^\dagger). \tag{1}$$

We group these operators into a vector

$$\hat{\mathbf{r}} = (\hat{q}_1, \hat{p}_1, ..., \hat{q}_n, \hat{p}_n)^T, \tag{2}$$

whose elements satisfy the following canonical commutation relations:

$$[\hat{r}_i, \hat{r}_j] = i\Omega_{ij}, \tag{3}$$

where $\Omega_{ij}$ is the matrix elements of a $2n \times 2n$ skew-symmetric matrix

$$\Omega = \bigoplus_{i=1}^{n} \begin{pmatrix} 0 & 1 \\ -1 & 0 \end{pmatrix}, \tag{4}$$

---

* These authors contributed equally to this work.
† wenweiho@nus.edu.sg

known as the $n$-mode symplectic form.

Given a bosonic state described by density matrix $\hat{\rho}$, all physical information is contained in its Wigner characteristic function $\chi$, defined as: $\chi(\boldsymbol{\xi}) = \text{Tr}(\hat{\rho}\hat{D}(\boldsymbol{\xi}))$, where $\hat{D}(\boldsymbol{\xi}) = e^{i\hat{\mathbf{r}}^T\Omega\boldsymbol{\xi}}$ is the Weyl operator and $\boldsymbol{\xi} \in \mathbb{R}^{2n}$. Indeed, the inverse transformation too is given by: $\hat{\rho} = \int \frac{d^{2n}\boldsymbol{\xi}}{(2\pi)^{2n}}\chi(\boldsymbol{\xi})\hat{D}(-\boldsymbol{\xi})$. Equivalently, all physical information is contained in its Wigner function too, which is defined as: $W(\mathbf{x}) = \frac{1}{\pi^n}\int_{\mathbb{R}^n}\langle \mathbf{q} - \mathbf{q}'|\hat{\rho}|\mathbf{q} + \mathbf{q}'\rangle e^{i2\mathbf{q}'\cdot\mathbf{p}}d^n\mathbf{q}'$, which is the Fourier transform of $\chi(\boldsymbol{\xi})$. Here $|\mathbf{q} \pm \mathbf{q}'\rangle$ are eigenstates of the position operators $\hat{\mathbf{x}}$ with continuous eigenvalues $\mathbf{q} \pm \mathbf{q}'$ with $\mathbf{q}, \mathbf{q}', \mathbf{p} \in \mathbb{R}^n$, and $\mathbf{x} = (q_1, p_1, ..., q_n, p_n)^T \in \mathbb{R}^{2n}$.

The Wigner representation ($\chi$ or $W$) can be characterized by their statistical moments. The first two moments, known as the displacement vector and the covariance matrix, are defined as: $\mathbf{r} := \langle\hat{\mathbf{r}}\rangle = \text{Tr}(\hat{\rho}\hat{\mathbf{r}})$ and $V_{ij} := \langle\{\hat{\mathbf{r}}_i - \mathbf{r}_i, \hat{\mathbf{r}}_j - \mathbf{r}_j\}\rangle$, where $\{,\}$ is the anti-commutator. We see $\mathbf{r}$ is a $2n$-dimensional real vector, and $V$ is a $2n \times 2n$ dimensional real symmetric positive definite matrix which must satisfy the uncertainty principle $V + i\Omega \geq 0$.

Bosonic Gaussian states are defined as quantum states whose Wigner representation (either the Wigner characteristic $\chi$ or Wigner function $W$) describes a multi-variate normal distribution:

$$\chi(\boldsymbol{\xi}) = e^{-\frac{1}{4}\boldsymbol{\xi}^T\Omega^T\mathbf{V}\Omega\boldsymbol{\xi} + i(\Omega\mathbf{r})^T\boldsymbol{\xi}}, \quad W(\mathbf{x}) = \frac{e^{-(\mathbf{x}-\mathbf{r})\mathbf{V}^{-1}(\mathbf{x}-\mathbf{r})}}{(2\pi)^n\sqrt{\det V}}, \tag{5}$$

that is, they are both fully determined by the displacement vector $\mathbf{r}$ and the covariance matrix $V$. Higher moments can be computed, as is standard, via Wick's theorem. Note in our work, the generator state used to construct the projected ensemble is always assumed, without loss of generality, to have zero displacement $\mathbf{r} = \mathbf{0}$.

### B. Gaussian unitaries

In this work, we focus exclusively on the manifold of Gaussian states. Therefore, we restrict our attention to unitary dynamics that preserve the Gaussian nature of states, known as Gaussian unitary dynamics. Now, a Gaussian unitary induces a transformation of the quadrature operators $\hat{\mathbf{r}}$ via an affine map $(S, \mathbf{d}) := \hat{\mathbf{r}} \to S\hat{\mathbf{r}} + \mathbf{d}$, where $S \in \text{Sp}(2n, \mathbb{R})$ is a $2n \times 2n$ real symplectic matrix and $\mathbf{d} \in \mathbb{R}^{2n}$. Consequently, the transformation of the statistical moments of Gaussian states is given by:

$$\mathbf{r} \to S\mathbf{r} + \mathbf{d}. \quad V \to SVS^T. \tag{6}$$

Conversely, every pair $(S, \mathbf{d})$ of symplectic transformations and displacements acting on the phase space corresponds to some Gaussian unitary acting on the Hilbert space. Specifically, a Gaussian unitary is called passive when it preserves the particle number density $\nu = \langle\sum_{i=1}^n \hat{a}_i^\dagger\hat{a}_i\rangle/n = (\text{Tr}(V)/2n - 1)/2$.

Note that every *pure* Gaussian state can be obtained by performing some Gaussian unitary on a vacuum state (which has $\mathbf{r}_{vac} = 0$ and $V_{vac} = \mathbb{I}_{2n}$). It thus follows that the covariance matrix of a general pure Gaussian state can be written as: $V = SS^T$, where $S \in \text{Sp}(2n, \mathbb{R})$ is determined by the Gaussian unitary.

### C. Gaussian measurements

In our work, we consider a bipartite CV system composed of a $k$-mode subsystem $A$ and an $n - k$ mode subsystem $B$, and construct the projected ensemble on $A$. This involves measurements on $B$ and conditional updates on $A$. We describe here the formalism to understand its construction. Without loss of generality, we consider an initial Gaussian state on $n$ modes with zero displacement and covariance matrix $V$, which can be written as

$$V = \begin{pmatrix} V_A & V_{AB} \\ V_{AB}^T & V_B \end{pmatrix}. \tag{7}$$

Here, $V_A$ is the top-left $2k \times 2k$ submatrix, $V_B$ is the bottom right $2(n-k) \times 2(n-k)$ submatrix, and $V_{AB} = V_{AB}^T$ is a $2k \times 2(n-k)$ submatrix. The decomposition reflects correlations within $A(B)$, captured by $V_{A(B)}$, and correlations between, captured by $V_{AB}$. Note that $V_A$ and $V_B$ are the covariance matrices of the reduced density matrices on $A$ and $B$ respectively, on the first $k$ and last $n - k$ modes respectively.

Next, we consider performing a Gaussian measurement on subsystem $B$. This is specified by a positive operator-valued measure (POVM) specified by a set of rank-1 Gaussian state projectors

$$\{\hat{\Pi}(\mathbf{r}_B, \sigma_B) = |\phi(\mathbf{r}_B, \sigma_B)\rangle\langle\phi(\mathbf{r}_B, \sigma_B)|/(2\pi)^{n-k}\}_{\mathbf{r}_B}, \tag{8}$$

where $|\phi(\mathbf{r}_B, \sigma_B)\rangle$ is a bosonic Gaussian state with displacement $\mathbf{r}_B$ and some fixed, common covariance matrix $\sigma_B$. The choice of covariance matrix $\sigma_B$ determines the basis of measurement. Note that

$$\int d\mathbf{r}_B \hat{\Pi}(\mathbf{r}_B, \sigma_B) = \hat{I}_B \tag{9}$$

for any fixed $\sigma_B$. To see this, consider first the special choice of coherent-state (heterodyne) measurements $\sigma_B = \mathbb{I}_B$, so that the projector $\hat{\Pi}(\mathbf{r}_B, \mathbb{I}_B) = |\mathbf{r}_B\rangle\langle\mathbf{r}_B|/(2\pi)^{n-k}$, where $|\mathbf{r}_B\rangle$ is an unsqueezed coherent state with displacement $\mathbf{r}_B$. It is well-known that the coherent states form an overcomplete basis

$$\int \frac{d\mathbf{r}_B}{(2\pi)^{n-k}} |\mathbf{r}_B\rangle\langle\mathbf{r}_B| = \hat{I}_B. \tag{10}$$

Next, we move to more general-dyne measurements representing measurements on $B$ in general Gaussian states basis [2]. The corresponding projectors $\hat{\Pi}(\mathbf{r}_B, \sigma_B)$ are then a Gaussian state with a covariance matrix $\sigma_B = SS^T$ and displacement $\mathbf{r}_B \in \mathbb{R}^{2n-2k}$, where $S$ is a general $2(n-k) \times 2(n-k)$ real symplectic matrix. It follows that

$$\begin{aligned}
\int d\mathbf{r}_B \hat{\Pi}(\mathbf{r}_B, \sigma_B) &= \int \frac{d\mathbf{r}_B}{(2\pi)^{n-k}} \hat{D}(\mathbf{r}_B)\hat{U}_S|0\rangle\langle 0|\hat{U}_S^\dagger \hat{D}(-\mathbf{r}_B) \\
&= \int \frac{d\mathbf{r}_B}{(2\pi)^{n-k}} \hat{D}(S\mathbf{r}_B)\hat{U}_S|0\rangle\langle 0|\hat{U}_S^\dagger \hat{D}(-S\mathbf{r}_B) \\
&= \int \frac{d\mathbf{r}_B}{(2\pi)^{n-k}} \hat{U}_S\hat{D}(\mathbf{r}_B)|0\rangle\langle 0|\hat{D}(-\mathbf{r}_B)\hat{U}_S^\dagger \\
&= \int \frac{d\mathbf{r}_B}{(2\pi)^{n-k}} \hat{U}_S|\mathbf{r}_B\rangle\langle\mathbf{r}_B|\hat{U}_S^\dagger \\
&= \hat{I}_B,
\end{aligned} \tag{11}$$

where $\hat{D}(\mathbf{r}_B)$ is the Weyl operator, $\hat{U}_S$ is a Gaussian unitary on subsystem B, which corresponds to the symplectic matrix $S$. In the second step, we changed the integration variable and used the fact that $\det(S) = 1$ when $S$ is a symplectic matrix. Note that in our work, we focus only on Gaussian measurements in product squeezed states basis, such that the Gaussian unitary is the squeezing operator $\otimes_{i=1}^{n-k} \exp\left(s_{\sigma,i}(\hat{a}_i^2 - \hat{a}_i^{\dagger 2})/2\right)$ with squeezing parameter $\mathbf{s}_\sigma = (s_{\sigma,1}, ..., s_{\sigma,n-k})$, and the corresponding covariance matrix $\sigma_B = \bigoplus_{i=1}^{n-k} \mathrm{diag}(e^{2s_{\sigma,i}}, e^{-2s_{\sigma,i}})$.

Under a Gaussian measurement $\{\hat{\Pi}(\mathbf{r}_B, \sigma_B)\}_{\mathbf{r}_B}$ on a Gaussian initial state with zero displacement, we will get a measurement outcome $\mathbf{r}_B$ (i.e., the displacement on $B$) and a projected state $|\psi(\mathbf{r}_A(\mathbf{r}_B), \widetilde{V}_A)\rangle$ on subsystem $A$, which is again a bosonic Gaussian state, with displacement $\mathbf{r}_A$ and covariance $\widetilde{V}_A$. Using the Wigner representation Eq. (5), this can be computed as follows:

$$\begin{aligned}
|\psi(\mathbf{r}_A(\mathbf{r}_B), \widetilde{V}_A)\rangle\langle\psi(\mathbf{r}_A(\mathbf{r}_B), \widetilde{V}_A)| &\propto \mathrm{Tr}_B(\hat{\rho}|\phi(\boldsymbol{r}_B, \sigma_B)\rangle\langle\phi(\boldsymbol{r}_B, \sigma_B)|) \\
&= \int \frac{d^{2n}\boldsymbol{\xi}}{(2\pi)^{2n}} \chi_\rho(\boldsymbol{\xi})\mathrm{Tr}_B(\hat{D}(-\boldsymbol{\xi})|\phi(\boldsymbol{r}_B, \sigma_B)\rangle\langle\phi(\boldsymbol{r}_B, \sigma_B)|) \\
&\propto \int \frac{d^{2k}\boldsymbol{\xi}_A}{(2\pi)^{2k}} e^{-\frac{1}{4}\boldsymbol{\xi}_A^T(V_A - V_{AB}(V_B + \sigma_B)^{-1}V_{AB}^T)\boldsymbol{\xi}_A + i\boldsymbol{\xi}_A^T V_{AB}(V_B + \sigma_B)^{-1}\mathbf{r}_B},
\end{aligned} \tag{12}$$

where $\boldsymbol{\xi} = (\boldsymbol{\xi}_A^T, \boldsymbol{\xi}_B^T)^T \in \mathbb{R}^{2n}$, so

$$\mathbf{r}_A = V_{AB}(V_B + \sigma_B)^{-1}\mathbf{r}_B, \tag{13}$$

$$\widetilde{V}_A = V_A - V_{AB}(V_B + \sigma_B)^{-1}V_{AB}^T. \tag{14}$$

We see that the displacement $\mathbf{r}_A$ depends on the measurement outcome $\mathbf{r}_B$ but not the covariance matrix $\widetilde{V}_A$. The calculation also yields the corresponding Born probability of this outcome happening: the probability density $p(\mathbf{r}_B)$ is a Gaussian

$$p(\mathbf{r}_B) = \frac{e^{-\mathbf{r}_B^T(V_B + \sigma_B)^{-1}\mathbf{r}_B}}{\pi^{n-k}\sqrt{\mathrm{Det}(V_B + \sigma_B)}}. \tag{15}$$

It follows then that

$$p(\mathbf{r}_A) = \frac{e^{-\frac{1}{2}\mathbf{r}_A^T\Sigma_A^{-1}\mathbf{r}_A}}{(2\pi)^k\sqrt{\mathrm{Det}(\Sigma_A)}}, \tag{16}$$

where $\Sigma_A = (V_A - \widetilde{V}_A)/2$.

## II. RANDOM PURE BOSONIC GAUSSIAN STATES

In this section, we define the class of random pure bosonic Gaussian states analyzed in the main text. Our setup and notations are similar to models defined in [3, 4].

### A. Definition

For simplicity, in this section we consider a rearrangement of the quadrature operators as

$$\hat{\mathbf{r}} = (\hat{q}_1, ..., \hat{q}_n, \hat{p}_1, ..., \hat{p}_n)^T, \tag{17}$$

which only differs from that in Eq. (2) of Section I by a change of basis, specifically, a permutation. The matrix representation of the displacement vector and the covariance matrix then change correspondingly. Consider now the covariance matrix $V$ of a (zero-displacement) pure bosonic Gaussian state. As mentioned, it is decomposable into $V = SS^T$, where $S$ is a real symplectic matrix. We thus define random pure bosonic Gaussian states by randomly sampling $S$ from the symplectic group $\mathrm{Sp}(2n, \mathbb{R})$. However, owing to the non-compactness of this group, there is no natural way to sample from this space in a canonical fashion, such as in a uniform way. To overcome this limitation, we examine the Bloch-Messiah decomposition of a real symplectic matrix [5, 6], which states that $S$ can be decomposed as follows:

$$S = O_1[D \oplus D^{-1}]O_2, \tag{18}$$

where $D$ is an $n$-dimensional diagonal non-negative real matrix, and $O_1, O_2$ are real symplectic and orthogonal matrices, i.e., $O_1, O_2 \in \mathrm{Sp}(2n, \mathbb{R}) \cap O(2n, \mathbb{R})$. It turns out that the latter space is compact, and in fact isomorphic to the complex unitary group $\mathrm{Sp}(2n, \mathbb{R}) \cap O(2n, \mathbb{R}) \cong \mathrm{U}(n)$, the complex unitary group of $n$-matrices, which *does* possess a uniform, normalizable, Haar measure. Using this, we see that the covariance matrix of a pure bosonic Gaussian state can be written

$$V = SS^T = O_1[D^2 \oplus D^{-2}]O_1^T. \tag{19}$$

Physically, the diagonal matrices $D$ represent the action of a tensor product of one-mode squeezing operations, which changes the particle-number of the system, while $O_1$ represents the action of a potentially entangling, passive (number-conserving) unitary. We see the origin of the lack of compactness of the space of symplectic matrices: active unitaries can change the particle-number of the system, which need not be bounded from above. Conversely, if we *fix* the particle-number of the system, then the set of unitary operations that preserve this number does form a compact group. This thus motivates us to define our random pure bosonic Gaussian states as follows: define an initial state which is a tensor product of 1-mode squeezed states with a uniform squeezing strength $s \in \mathbb{R}$ (the assumption of uniformity can of course be relaxed) and zero displacement, which has covariance matrix

$$V_0 = D^2 \oplus D^{-2} = \begin{pmatrix} e^{2s}\mathbb{1}_n & \\ & e^{-2s}\mathbb{1}_n \end{pmatrix}. \tag{20}$$

Then, randomly apply a passive (particle-conserving) unitary $U$ transformation from the uniform Haar distribution (on the unitary group) to the initial squeezed state to obtain $|\Psi\rangle$. In phase space, this transformation corresponds to the following symplectic-ortho matrix:

$$O(U) = \begin{pmatrix} \mathrm{Re}(U) & \mathrm{Im}(U) \\ -\mathrm{Im}(U) & \mathrm{Re}(U) \end{pmatrix}. \tag{21}$$

In other words, the random bosonic Gaussian states we consider in the main text are defined as *uniformly random states within the manifold of Gaussian states with fixed particle number*. As we are in fact considering a *family* of states (labeled by the system size $n$), we need also a way to specify how the particle number changes with $n$. In our work, we adopt the conceptually simplest scenario where the particle-number density $\nu$ is constant across system sizes.

To summarize, the random bosonic Gaussian states $|\Psi\rangle$ we consider in this work are parameterized by a squeezing parameter $s$ and have zero displacement and covariance

$$V = OV_0O^T = \cosh(2s)\mathbb{1}_{2n} + \sinh(2s)OYO^T, \tag{22}$$

where $Y = \mathbb{1}_n \oplus (-\mathbb{1}_n)$, and $O(U)$ is a $2n \times 2n$ real random ortho-symplectic matrix induced by an $n \times n$ complex Haar random unitary $U$. These states all have particle number density $\nu = \langle \sum_{i=1}^n \hat{a}_i^\dagger \hat{a}_i \rangle / n = (\mathrm{Tr}(V)/2n - 1)/2 = (\cosh(2s) - 1)/2$.

### B. Useful notation for averages over subsystems in random bosonic Gaussian states

In anticipation of our Theorems 1 and 2, it is useful to introduce some notation dealing with subsystems of the global system and their Haar averages. Specifically, we introduce in this subsection the isometries $P_A$, $P_B$ and projectors $\Pi_A, \Pi_B$ used to extract the submatrices $V_A, V_B$ and $V_{AB}$ from $V$; and we also demonstrate how to compute $\mathbb{E}[V_A]$.

Define the isometries

$$P_A := \begin{pmatrix} \mathbb{I}_k & 0_{k,n-k} \end{pmatrix} \oplus \begin{pmatrix} \mathbb{I}_k & 0_{k,n-k} \end{pmatrix}, \quad P_B := \begin{pmatrix} 0_{n-k,k} & \mathbb{I}_{n-k} \end{pmatrix} \oplus \begin{pmatrix} 0_{n-k,k} & \mathbb{I}_{n-k} \end{pmatrix}, \tag{23}$$

and projectors

$$\Pi_A := \pi_A \oplus \pi_A, \quad \Pi_B := \pi_B \oplus \pi_B, \tag{24}$$

where $\pi_A, \pi_B$ are defined as

$$\pi_A := \begin{pmatrix} \mathbb{I}_k & 0_{k,n-k} \\ 0_{n-k,k} & 0_{n-k,n-k} \end{pmatrix}, \quad \pi_B := \begin{pmatrix} 0_{k,k} & 0_{k,n-k} \\ 0_{n-k,k} & \mathbb{I}_{n-k} \end{pmatrix} = \mathbb{I}_n - \pi_A. \tag{25}$$

Then the submatrices of $V$, Eq. (7), can be expressed

$$V_A = P_A V P_A^T, \quad V_B = P_B V P_B^T, \quad V_{AB} = P_A V P_B^T. \tag{26}$$

For the random bosonic Gaussian states we introduce, using Eq. (21) and Eq. (22), we have

$$V_A = \cosh(2s)\mathbb{I}_{2k} + \sinh(2s)P_A OYO^T P_A^T, \tag{27}$$

where

$$P_A OYO^T P_A^T = \frac{1}{2} P_A \begin{pmatrix} (UU^T + U^*U^\dagger) & i(UU^T - U^*U^\dagger) \\ i(UU^T - U^*U^\dagger) & -(UU^T + U^*U^\dagger) \end{pmatrix} P_A^T. \tag{28}$$

We desire to understand $\mathbb{E}[V_A]$, which is the covariance matrix of the averaged density matrix on $A$. Using Weingarten calculus [7, 8] we can easily compute the expectation value over the Haar measure as

$$\begin{aligned} \mathbb{E}[V_A] &= \int_{U \sim \text{Haar}(2n)} \mathrm{d}U \, V_A \\ &= \int_{U \sim \text{Haar}(2n)} \mathrm{d}U \cosh(2s)\mathbb{I}_{2k} \\ &= \cosh(2s)\mathbb{I}_A, \end{aligned} \tag{29}$$

using that $\int_{\text{Haar}} \mathrm{d}U U U^T = \int_{\text{Haar}} \mathrm{d}U U^*U^\dagger = 0$.

## III. PROOF OF THEOREM 1

In this section, we provide a detailed proof of Theorem 1 in the main text, which states that in the TDL, the projected state of the PE are all unsqueezed coherent states on A with unit probability. For convenience, we restate Theorem 1:

**Theorem 1.** *The (common) covariance matrix $\widetilde{V}_A$ of the projected states $|\psi(\mathbf{r}_A, \widetilde{V}_A)\rangle$ on $k$-modes, generated from heterodyne measurements on the complement of a random $n$-mode BGS, obeys for any $\epsilon > 0$*

$$\mathbb{P}(\|\widetilde{V}_A - \mathbb{I}_A\|_1 \geq \epsilon) \leq C(1 + \epsilon/(2k))/\epsilon^2 n, \tag{30}$$

*where $C$ is a constant depending on $k, s$ but not on $n$. Here $\|\cdot\|_1$ is the trace norm.*

To prove Theorem 1, we first introduce the following Lemma to bound the probability:

**Lemma 1.** *The (common) covariance matrix $\widetilde{V}_A$ of the projected states $|\psi(\mathbf{r}_A, \widetilde{V}_A)\rangle$ on $k$-modes, generated from heterodyne measurements on the complement of a random $n$-mode BGS, obeys for any $\epsilon > 0$*

$$\mathbb{P}(\|\widetilde{V}_A - \mathbb{I}_A\|_1 \geq \epsilon) \leq \frac{4k^2(1 + \epsilon/2k)}{\epsilon^2} \mathbb{E}[\text{Tr}(\widetilde{V}_A - \mathbb{I}_A)] \tag{31}$$

*where $\mathbb{E}[\cdot]$ denotes the expectation value over the Haar measure.*

*Proof.* Since the conditional state is pure, by considering the Bloch-Messiah decomposition of the positive, symplectic covariance matrix $\widetilde{V}_A$ we know that its spectrum consists of eigenvalues that come in pairs $\{x_i, x_i^{-1}\}_{i=1}^k$. We can assume without loss of generality $x_i \geq 1$. The trace distance of $\widetilde{V}_A$ to the identity has then a simple closed form expression in terms of the spectrum:

$$\|\widetilde{V}_A - \mathbb{I}_A\|_1 = \sum_{i=1}^k x_i - x_i^{-1}. \tag{32}$$

This comes from observing that the singular values of the argument are $\{(x_i - 1), (1 - x_i^{-1})\}_{i=1}^k$. On the other hand, $\text{Tr}(\widetilde{V}_A - I_A)$, has the explicit expression in terms of the eigenvalues

$$\text{Tr}(\widetilde{V}_A - I_A) = \sum_{i=1}^k (x_i + x_i^{-1} - 2) = \sum_{i=1}^k \frac{(x_i - 1)^2}{x_i}. \tag{33}$$

### 1. 1-dimensional base case

We first study the base case $k = 1$. Let $\epsilon > 0$ and suppose $x - x^{-1} \geq \epsilon$. This is equivalent to

$$x \geq \frac{1}{2}(\epsilon + \sqrt{4 + \epsilon^2}). \tag{34}$$

Now,

$$\frac{1}{2}(\epsilon + \sqrt{4 + \epsilon^2}) \geq \frac{1}{2}(\epsilon + 2) = \frac{\epsilon}{2} + 1. \tag{35}$$

Therefore,

$$\mathbb{P}(x - x^{-1} \geq \epsilon) \leq \mathbb{P}(x \geq \epsilon/2 + 1) = \mathbb{P}(x - 1 \geq \epsilon/2). \tag{36}$$

Now, consider $x - 1 \geq \epsilon/2$. We have

$$\frac{(x - 1)^2}{x} \geq \frac{(\epsilon/2)^2}{1 + \epsilon/2}. \tag{37}$$

This inequality holds because the function $(x - 1)^2/x$ is strictly increasing (in the domain $x \geq 1$), so plugging in $x = 1 + \epsilon/2$ gives us a lower bound, i.e., the RHS.

Next, let $A$ denote the event that $x - 1 \geq \epsilon/2$. We have

$$\mathbb{E}\left[\frac{(x - 1)^2}{x}\right] \geq \mathbb{E}\left[\mathbb{1}_A \frac{(x - 1)^2}{x}\right] \geq \frac{(\epsilon/2)^2}{1 + \epsilon/2} \mathbb{P}(A), \tag{38}$$

where $\mathbb{1}_A$ is indicator function for event $A$, and in the last inequality we lower bound $(x - 1)^2/x$ by its minimum value in $A$. Rearranging, we get

$$\mathbb{P}(x - 1 \geq \epsilon/2) \leq \frac{1 + \epsilon/2}{(\epsilon/2)^2} \mathbb{E}\left[\frac{(x - 1)^2}{x}\right]. \tag{39}$$

Therefore, we have shown for the $k = 1$ case

$$\mathbb{P}(x - x^{-1} \geq \epsilon) \leq \frac{1 + \epsilon/2}{(\epsilon/2)^2} \mathbb{E}\left[\frac{(x - 1)^2}{x}\right]. \tag{40}$$

### 2. Higher-dimensions

Next, consider the higher dimensions $k \geq 2$. We introduce the non-negative vector

$$\boldsymbol{z} = (z_1, z_2, \cdots, z_k) \tag{41}$$

where $z_i = (x_i - x_i^{-1})$. We consider vector $p$-norms for any $p \in [1, \infty]$:

$$\|\boldsymbol{v}\|_p := \left( \sum_{i=1}^{k} |v_i|^p \right)^{1/p}. \tag{42}$$

In the context of our particular vector $\boldsymbol{z}$, because each entry is non-negative,

$$\|\boldsymbol{z}\|_1 = \sum_{i=1}^{k} (x_i - x_i^{-1}) \tag{43}$$

which is the desired trace distance $\|\widetilde{V}_A - \mathbb{I}_A\|_1$. We will work with

$$\|\boldsymbol{z}\|_\infty = \max_i (x_i - x_i^{-1}). \tag{44}$$

Now, we note $\|\boldsymbol{z}\|_\infty \geq \epsilon$ if and only if $z_i \geq \epsilon$ for some $i$. Therefore

$$
\begin{aligned}
\mathbb{P}(\|\boldsymbol{z}\|_\infty \geq \epsilon) &\leq \sum_{i=1}^{k} \mathbb{P}(x_i - x_i^{-1} \geq \epsilon) \\
&< \sum_{i=1}^{k} \frac{1 + \epsilon/2}{(\epsilon/2)^2} \mathbb{E}\left[ \frac{(x_i - 1)^2}{x_i} \right] \\
&= \frac{1 + \epsilon/2}{(\epsilon/2)^2} \mathbb{E}\left[ \sum_{i=1}^{k} \frac{(x_i - 1)^2}{x_i} \right] \\
&= \frac{1 + \epsilon/2}{(\epsilon/2)^2} \mathbb{E}\left[ \mathrm{Tr}(\widetilde{V}_A - I_A) \right].
\end{aligned}
\tag{45}
$$

Lastly, we note that all norms are equivalent and the inequality $\|\boldsymbol{z}\|_\infty \geq \|\boldsymbol{z}\|_1/k$ holds. Thus, we have

$$\mathbb{P}(\|\boldsymbol{z}\|_1 \geq \epsilon) \leq \mathbb{P}(\|\boldsymbol{z}\|_\infty \geq \epsilon/k) \leq 4k^2 \frac{1 + \epsilon/(2k)}{\epsilon^2} \mathbb{E}[\mathrm{Tr}(\widetilde{V}_A - \mathbb{I}_A)]. \tag{46}$$

This concludes the proof of Lemma 1. ∎

*Proof of Theorem 1.* To prove Theorem 1, we use Lemma 1 and upper bound $\mathbb{E}[\mathrm{Tr}(\widetilde{V}_A - \mathbb{I}_A)]$ by explicit calculation. Consider the expression of the (common) covariance matrix $\widetilde{V}_A$,

$$\widetilde{V}_A = V_A - V_{AB}(V_B + \mathbb{I}_B)^{-1} V_{AB}^T. \tag{47}$$

In Section II, we have obtained the expectation value of the first term in Eq. (47), so its trace is

$$\mathbb{E}[\mathrm{Tr}(V_A)] = 2k \cosh(2s). \tag{48}$$

Therefore we need only compute the expectation value of $\mathrm{Tr}(V_{AB}(V_B + \mathbb{I}_B)^{-1} V_{AB}^T)$. Consider

$$V_B + \mathbb{I}_B = (1 + \cosh(2s))\mathbb{I}_B + \sinh(2s)P_B O Y O^T P_B^T, \tag{49}$$

where $Y = \mathbb{I}_n \oplus (-\mathbb{I}_n)$, $O$ is a random symplectic-orthogonal matrix defined in Eq. (21), and $P_B$ is the matrix defined in Eq. (23). Note that the second term is the principal submatrix of a real Hermitian matrix $O Y O^T$, whose eigenvalues are $\pm 1$. Thus, we can conclude by the Poincaré separation theorem or Cauchy interlacing theorem that the eigenvalues of $P_B O Y O^T P_B^T$, denoted as $y_i$, are bounded as

$$-1 \leq y_i \leq 1. \tag{50}$$

Using this and noting that for any $s \in \mathbb{R}$, since $1 + \cosh(2s) > |\sinh(2s)|$, we can expand the term $(V_B + \mathbb{I}_B)^{-1}$ as an absolutely convergent Taylor series

$$(V_B + \mathbb{I}_B)^{-1} = \frac{1}{(1 + \cosh(2s))} \sum_{m=0}^{\infty} (-u)^m (P_B O Y O^T P_B^T)^m, \tag{51}$$

where $u := \sinh(2s)/(1 + \cosh(2s))$ which satisfies $-1 < u < 1$. We have

$$\begin{aligned}
&\text{Tr}(V_{AB}(V_B + \mathbb{I}_B)^{-1} V_{AB}^T) \\
&= \frac{1}{(1 + \cosh(2s))} \text{Tr}(\Pi_A V \Pi_B \sum_{m=0}^{\infty} (-u)^m (\Pi_B O Y O^T \Pi_B)^m \Pi_B V \Pi_A) \\
&= \frac{\sinh^2(2s)}{(1 + \cosh(2s))} \sum_{m=0}^{\infty} (-u)^m \text{Tr}(\Pi_A O Y O^T \Pi_B (\Pi_B O Y O^T \Pi_B)^m \Pi_B O Y O^T \Pi_A),
\end{aligned} \tag{52}$$

where $\Pi_A, \Pi_B$ are projctors defined in Eq. (24). Since we only consider the trace, we can use projectors $\Pi_A, \Pi_B$ to replace $P_A, P_B$. In the last equality, we use the fact that $\Pi_A \Pi_B = 0$.

Define the matrices

$$A_j := \Pi_A O Y O^T \Pi_B X^j \Pi_B O Y O^T \Pi_A, \tag{53}$$

where $X = \Pi_B O Y O^T \Pi_B$. Since the power series converges absolutely, we can exchange the sums to get

$$\mathbb{E}[\text{Tr}(V_{AB}(V_B + I_B)^{-1} V_{AB}^T)] = \frac{\sinh(2s)^2}{(1 + \cosh(2s))} \sum_{m=0}^{\infty} (-u)^m \mathbb{E}[\text{Tr}(A_m)]. \tag{54}$$

Therefore, we only need to compute $\mathbb{E}[\text{Tr}(A_m)]$. We first consider the odd terms $\mathbb{E}[\text{Tr}(A_{2j+1})]$. We observe that matrix $X = \Pi_B O Y O^T \Pi_B$ can be written as

$$X = \begin{pmatrix} a_0 & b_0 \\ b_0 & -a_0 \end{pmatrix}, \tag{55}$$

where $a_0, b_0$ are $n - k \times n - k$ dimensional matrices. Then, employing induction and matrix multiplication, we can establish a similar expression for $A_{2j+1}$:

$$\begin{aligned}
A_{2j+1} &= \Pi_A O Y O^T \Pi_B X^{2j+1} \Pi_B O Y O^T \Pi_A \\
&= \begin{pmatrix} c_j & d_j \\ d_j & -c_j \end{pmatrix},
\end{aligned} \tag{56}$$

where $c_j$ and $d_j$ represent $n - k \times n - k$ dimensional matrices. Consequently, we deduce the vanishing of odd terms:

$$\mathbb{E}[\text{Tr}(A_{2j+1})] = 0. \tag{57}$$

Next, we consider the even terms. we have

$$\begin{aligned}
\text{Tr}(A_{2j}) &= \text{Tr}((\mathbb{I}_B - \Pi_B) O Y O^T \Pi_B (\Pi_B O Y O^T \Pi_B)^{2j} \Pi_B O Y O^T) \\
&= \text{Tr}(\Pi_B (\Pi_B O Y O^T \Pi_B)^{2j}) - \text{Tr}((\Pi_B O Y O^T \Pi_B)^{2j+2}) \\
&= \text{Tr}(X^{2j}) - \text{Tr}(X^{2j+2}) \\
&= \sum_i (y_i^{2j} - y_i^{2j+2}) \geq 0,
\end{aligned} \tag{58}$$

where $y_i$ denotes the eigenvalue of $P_B O Y O^T P_B^T$, satisfying the inequality $-1 \leq y_i \leq 1$. In the first line, we used that $\Pi_A = \mathbb{I}_{2n} - \Pi_B$; and in the last line, we used the fact that the non-zero eigenvalues of matrix $X$ are same as that of matrix $P_B O Y O^T P_B^T$.

We can show $\text{Tr}(A_{2j})$ decreases with $j$, viz

$$\begin{aligned}
\text{Tr}(A_{2j}) - \text{Tr}(A_{2j+2}) &= \sum_i ((\lambda_i^{2j} - y_i^{2j+2}) - (y_i^{2j+2} - y_i^{2j+4})) \\
&= \sum_i (y_i^{2j} + y_i^{2j+4} - 2y_i^{2j+2}) \geq 0.
\end{aligned} \tag{59}$$

Next, using the Weingarten calculus, we evaluate the second two terms,

$$\mathbb{E}[\mathrm{Tr}(A_0)] = 2k - \frac{2k(1+k)}{n+1}, \tag{60}$$

and

$$\mathbb{E}[\mathrm{Tr}(A_2)] = \frac{2k(1+k)(n+1-k)(n-k)}{n(n+1)(n+3)} < \frac{2k(1+k)}{n+3} < \frac{2k(1+k)}{n}. \tag{61}$$

Therefore in the TDL, for any $j \geq 1$ we have the following bound,

$$0 < \mathrm{Tr}(\mathbb{E}[A_{2j}]) < \frac{2k(1+k)}{n}. \tag{62}$$

This allows us to bound the expectation value of $\mathbb{E}[\mathrm{Tr}(\widetilde{V}_A - \mathbb{I}_A)]$ as

$$
\begin{aligned}
&\mathbb{E}[\mathrm{Tr}(\widetilde{V}_A - I_A)] = \mathbb{E}[\mathrm{Tr}(V_A)] - \mathbb{E}[\mathrm{Tr}(V_{AB}(V_B + I_B)^{-1}V_{AB}^T)] - 2k \\
&= 2k(\cosh(2s) - 1) - \frac{\sinh^2(2s)}{(1+\cosh(2s))}\left(\mathrm{Tr}(\mathbb{E}[A_0]) + \sum_{m=1}^{\infty}(-u)^{2m}\mathrm{Tr}(\mathbb{E}[A_{2m}])\right) \\
&\leq 2k(\cosh(2s) - 1) - \frac{\sinh^2(2s)}{(1+\cosh(2s))}\mathrm{Tr}(\mathbb{E}[A_0]) + \frac{\sinh^2(2s)}{(1+\cosh(2s))}\sum_{m=1}^{\infty}u^{2m}\mathrm{Tr}(\mathbb{E}[A_{2m}]) \\
&< 2k(\cosh(2s) - 1) - \frac{\sinh^2(2s)}{(1+\cosh(2s))}\left(2k - \frac{2k(1+k)}{n+1}\right) + \frac{\sinh^2(2s)}{(1+\cosh(2s))}\sum_{m=1}^{\infty}u^{2m}\frac{2k(1+k)}{n} \\
&< (\cosh(2s) - 1)\frac{2k(k+1)}{n}\left(1 + \sum_{m=1}^{\infty}u^{2m}\right) \\
&= (\cosh(2s) - 1)\left(\frac{2k(k+1)}{1-u^2}\right)\frac{1}{n},
\end{aligned}
\tag{63}
$$

where $u = \sinh(2s)/(1 + \cosh(2s))$. In the second line, we used the series expansion in Eq. (54); in the third line, we used the fact that $u^2 \geq 0$ and $\mathrm{Tr}(\mathbb{E}[A_{2m}]) > 0$; and in the fourth line, we used the bound of $\mathrm{Tr}(\mathbb{E}[A_{2m}])$.

Finally, by combining Lemma 1 and Eq. (63), we get

$$\mathbb{P}(\|\widetilde{V}_A - \mathbb{I}_A\|_1 \geq \epsilon) \leq 4k^2(1 + \epsilon/(2k))/\epsilon^2)\mathbb{E}[\mathrm{Tr}(\widetilde{V}_A - \mathbb{I}_A)] \leq C(1 + \epsilon/(2k)/\epsilon^2 n, \tag{64}$$

where $C = 4k^2(\cosh(2s) - 1)(\frac{2k(k+1)}{1-u^2})$ is a constant which we note only depends on $k, s$ but not $n$. $\quad\square$

## IV. PROOF OF THEOREM 2

In this section, we prove Theorem 2 of the main text, which establishes that with unit probability in the TDL, the distribution of displacements of the PE is statistically indistinguishable from those of an isotropic normal distribution with variance $\nu$. For convenience, we restate Theorem 2 here:

**Theorem 2.** *The distribution $\mathcal{N}_A = \mathcal{N}(\mathbf{0}, \Sigma_A)$ of displacements $\mathbf{r}_A$ of the projected states $|\psi(\mathbf{r}_A, \widetilde{V}_A)\rangle$ on $k$-modes, generated from heterodyne measurements on the complement of a random $n$-mode BGS, obeys for any $\epsilon > 0$*

$$\mathbb{P}(D_{\mathrm{KL}}(\mathcal{N}_A\|\mathcal{N}_\nu) \geq \epsilon) \leq C\frac{1 + \frac{\nu\epsilon'}{2k}}{\nu^2\epsilon'^2 n} + De^{-n\nu^2\epsilon'^2 F}, \tag{65}$$

*where $C, D, F$ are constants depending on $k, s$ only, and $\epsilon' = \frac{1}{2}\left(\sqrt{\epsilon/k}\sqrt{4 + \epsilon/k} - \epsilon/k\right)$. Above, $D_{\mathrm{KL}}$ refers to the Kullback-Leibler divergence [9] of $\mathcal{N}_A$ with respect to $\mathcal{N}_\nu = \mathcal{N}(\mathbf{0}, \nu\mathbb{I}_A)$. In full generality, the KL-divergence of distribution $p$ from $q$ is defined as*

$$D_{\mathrm{KL}}(p\|q) = \int \mathrm{d}\mathbf{r}\, p(\mathbf{r})\log\left(\frac{p(\mathbf{r})}{q(\mathbf{r})}\right), \tag{66}$$

*and for two d-dimensional multivariate normal distributions, denoted as $\mathcal{N}(\boldsymbol{\mu}_p, \Sigma_p)$ and $\mathcal{N}(\boldsymbol{\mu}_q, \Sigma_q)$, the KL divergence has a closed form expression,*

$$D_{\mathrm{KL}}(\mathcal{N}(\boldsymbol{\mu}_p, \Sigma_p)||\mathcal{N}(\boldsymbol{\mu}_q, \Sigma_q)) = \frac{1}{2}\left(\log\frac{\det(\Sigma_q)}{\det(\Sigma_p)} - d + (\boldsymbol{\mu}_p - \boldsymbol{\mu}_q)^T \Sigma_q^{-1}(\boldsymbol{\mu}_p - \boldsymbol{\mu}_q) + \mathrm{Tr}(\Sigma_q^{-1}\Sigma_p)\right). \tag{67}$$

*Thus, in our case*

$$D_{\mathrm{KL}}(\mathcal{N}_A||\mathcal{N}_\nu) = \frac{1}{2}\left(\log\frac{\det(\nu\mathbb{I}_A)}{\det(\Sigma_A)} - 2k + \mathrm{Tr}(\Sigma_A/\nu)\right). \tag{68}$$

As sketched in the main text, the proof of the theorem follows the following logic. We show (Lemma 2) that covariance matrix of the projected state $V_A$ is *concentrated* around its expected value $\mathbb{E}[V_A]$, as a result of Levy's lemma. Then we show (Lemma 3) that the covariance matrix $\Sigma_A$ governing the spread of displacements $\mathbf{r}_A$ is close to $\nu\mathbb{I}_A$, using that $\Sigma_A = (V_A - \widetilde{V}_A)/2$. Lastly, (Lemma 4) we related closeness of $\Sigma_A$ to $\nu\mathbb{I}_A$ to smallness of the KL divergence itself. To that end, we introduce the lemmas.

**Lemma 2.** *The covariance matrix $V_A$ of the reduced density matrix on $A$, generated by a random $n$-mode BGS, obeys for any $\epsilon > 0$*

$$\mathbb{P}(\|V_A - \mathbb{E}[V_A]\|_1 \geq \epsilon) \leq 8k^2 \exp\left(-\frac{n\epsilon^2}{1024e^{4s}k^2}\right). \tag{69}$$

*Proof.* We desire to prove that with high probability, $V_A$ *concentrates* around $\mathbb{E}[V_A] = \cosh(2s)\mathbb{I}_A$ in the thermodynamic limit. We first consider the following sequence of inequalities:

$$\begin{aligned}
\mathbb{P}(\|V_A - \mathbb{E}[V_A]\|_1 \geq \epsilon) &\leq \mathbb{P}(\|V_A - \mathbb{E}[V_A]\|_{\mathrm{Entry\text{-}wise},\ 1} \geq \epsilon) \\
&\leq \mathbb{P}(|(V_A - \mathbb{E}[V_A])_{ij}| \geq \epsilon/4k^2 \text{ for some } i, j) \\
&\leq 4k^2 \mathbb{P}(|(V_A - \mathbb{E}[V_A])_{ij}| \geq \epsilon/4k^2).
\end{aligned} \tag{70}$$

In the first line, we used that $\|A\|_1 \leq \|A\|_{\mathrm{entry\text{-}wise},1} := \sum_{ij}|A_{ij}|$; in the second line, we used that $\sum_{i=1}^{2k}\sum_{j=1}^{2k}|A_{ij}| > \epsilon \implies |A_{ij}| > \epsilon/4k^2$ for some $i, j$; and in the third line, we used the union bound.

Note that $(V_A)_{ij}(U)$ is a function of the $n \times n$ Haar random unitary $U$, and is Lipschitz continuous with Lipschitz constant $4e^{2s}$:

$$\begin{aligned}
|(V_A)_{ij} - (V'_A)_{ij}| &\leq \|V_A - V'_A\|_2 \\
&\leq \|V - V'\|_2 = \|O(U)V_0O^T(U) - O(U')V_0O^T(U')\|_2 \\
&\leq \|O(U)V_0(O^T(U) - O^T(U'))\|_2 + \|(O(U) - O(U'))V_0O^T(U')\|_2 \\
&\leq 2\|V_0\|_\infty \|(O(U) - O(U'))\|_2 \\
&\leq 4e^{2s}\|U - U'\|_2.
\end{aligned} \tag{71}$$

In the second inequality we used that $V_A = P_A V P_A^T$, and that the projection cannot increase norm. In the third inequality, we used

$$\|ABC\|_p \leq \|A\|_\infty \|B\|_\infty \|C\|_p, \tag{72}$$

for $p \in [1, \infty)$. The last inequality comes from

$$O(U) = \begin{pmatrix} \mathrm{Re}(U) & \mathrm{Im}(U) \\ -\mathrm{Im}(U) & \mathrm{Re}(U) \end{pmatrix} = F\begin{pmatrix} U & \\ & U^* \end{pmatrix}F^{-1}, \quad F = \frac{1}{\sqrt{2}}\begin{pmatrix} \mathbb{I}_n & i\mathbb{I}_n \\ i\mathbb{I}_n & \mathbb{I}_n \end{pmatrix}. \tag{73}$$

Therefore using Levy's lemma, Corollary 4.4.28 of [10], we have

$$\mathbb{P}(|(V_A - \mathbb{E}[V_A])_{ij})| \geq \delta) \leq 2\exp\left(-\frac{n\delta^2}{64e^{4s}}\right). \tag{74}$$

Putting it all together, we have for any $\epsilon > 0$,

$$\mathbb{P}(\|V_A - \mathbb{E}[V_A]\|_1 \geq \epsilon) \leq 8k^2 \exp\left(-\frac{n\epsilon^2}{1024e^{4s}k^2}\right). \tag{75}$$

This concludes the proof of Lemma 2. ∎

**Lemma 3.** *Let $\mathcal{N}_A = \mathcal{N}(\mathbf{0}, \Sigma_A)$ be the distribution of displacements $\mathbf{r}_A$ of the projected states $|\psi(\mathbf{r}_A, \widetilde{V}_A)\rangle$ on $k$-modes, generated from coherent-state measurements on the complement of a random $n$-mode BGS. Its variance $\Sigma_A$ obeys for any $\epsilon > 0$*

$$\mathbb{P}(\|\Sigma_A - \nu \mathbb{I}_A\|_1 \geq \epsilon) \leq C(1 + \epsilon/(2k))/\epsilon^2 n + 8k^2 \exp\left(-\frac{n\epsilon^2}{1024 e^{4s} k^2}\right). \tag{76}$$

*where $C$ is a constant depending on $k, s$ but not $n$. Here $\|\cdot\|_1$ is the trace norm.*

*Proof.* We first note that

$$\Sigma_A = (V_A - \widetilde{V}_A)/2. \tag{77}$$

Then, using the triangle inequality of the trace norm, we have:

$$\|\Sigma_A - \nu \mathbb{I}_A\|_1 \leq \frac{1}{2}\|V_A - (2\nu + 1)\mathbb{I}_A\|_1 + \frac{1}{2}\|\widetilde{V}_A - \mathbb{I}_A\|_1. \tag{78}$$

Therefore, we can bound the probability as follows:

$$\begin{aligned}
\mathbb{P}(\|\Sigma_A - \nu \mathbb{I}_A\|_1 \geq \epsilon) &\leq \mathbb{P}\left(\frac{1}{2}\|V_A - (2\nu + 1)\mathbb{I}_A\|_1 + \frac{1}{2}\|\widetilde{V}_A - \mathbb{I}_A\|_1 \geq \epsilon\right) \\
&\leq \mathbb{P}\left(\frac{1}{2}\|V_A - (2\nu + 1)\mathbb{I}_A\|_1 \geq \frac{\epsilon}{2} \cup \frac{1}{2}\|\widetilde{V}_A - \mathbb{I}_A\|_1 \geq \frac{\epsilon}{2}\right) \\
&\leq \mathbb{P}\left(\frac{1}{2}\|V_A - (2\nu + 1)\mathbb{I}_A\|_1 \geq \frac{\epsilon}{2}\right) + \mathbb{P}\left(\frac{1}{2}\|\widetilde{V}_A - \mathbb{I}_A\|_1 \geq \frac{\epsilon}{2}\right) \\
&\leq C(1 + \epsilon/(2k))/\epsilon^2 n + 8k^2 \exp\left(-\frac{n\epsilon^2}{1024 e^{4s} k^2}\right).
\end{aligned} \tag{79}$$

In the last inequality, we used Theorem 1 and Lemma 2. ∎

**Lemma 4.** *The normal distribution $\mathcal{N}(\mathbf{0}, \Sigma_A)$ obeys for any $\epsilon > 0$*

$$\mathbb{P}(D_{\mathrm{KL}}(\mathcal{N}(\mathbf{0}, \Sigma_A)||\mathcal{N}(\mathbf{0}, \nu \mathbb{I}_A)) \geq \epsilon) \leq \mathbb{P}(\|\Sigma_A - \nu \mathbb{I}_A\|_1 \geq \nu \epsilon'(\epsilon, k)) \tag{80}$$

*where $\epsilon' = \frac{1}{2}\left(\sqrt{\epsilon/k}\sqrt{4 + \epsilon/k} - \epsilon/k\right)$.*

*Proof.* First we consider the explicit expression for the KL divergence of $\mathcal{N}(\mathbf{0}, \Sigma_A)$ with respect to $\mathcal{N}(\mathbf{0}, \nu \mathbb{I}_A)$:

$$\begin{aligned}
D_{\mathrm{KL}}(\mathcal{N}(\mathbf{0}, \Sigma_A)||\mathcal{N}(\mathbf{0}, \nu \mathbb{I}_A)) &= \frac{1}{2}\left(-\log\det(\Sigma_A/\nu) + \mathrm{Tr}(\Sigma_A/\nu) - 2k\right) \\
&= \sum_{i=1}^{2k} \frac{1}{2}\left(-\log \lambda_i + \lambda_i - 1\right),
\end{aligned} \tag{81}$$

where $\lambda_i$ are the non-negative eigenvalues of $\Sigma_A/\nu$. Note each term in the sum is non-negative. Next we note the condition that

$$\sum_{i=1}^{2k} \frac{1}{2}\left(-\log \lambda_i + \lambda_i - 1\right) \leq \epsilon \tag{82}$$

implies

$$\frac{1}{2}\left(-\log \lambda_i + \lambda_i - 1\right) \leq \frac{\epsilon}{2k} \tag{83}$$

for *at least* one $i$ (because if this were not true for all $i$, their sum would be $< \epsilon$, a contradiction of the premise). Now because

$$\frac{1}{\lambda_i} + \lambda_i - 2 \geq -\log \lambda_i + \lambda_i - 1 \tag{84}$$

at least for one $i$ we have that

$$|\lambda_i - 1| \geq \frac{1}{2}\left(\sqrt{\epsilon/k}\sqrt{4+\epsilon/k} - \epsilon/k\right) \equiv \epsilon'(\epsilon, k), \tag{85}$$

which further implies

$$\|\Sigma_A - \nu\mathbb{I}_A\|_1 = \sum_i \nu|\lambda_i - 1| > \nu\epsilon'. \tag{86}$$

We therefore have that

$$\mathbb{P}(D_{\mathrm{KL}}(\mathcal{N}(\mathbf{0}, \Sigma_A)||\mathcal{N}(\mathbf{0}, \nu\mathbb{I}_A)) \geq \epsilon) \leq \mathbb{P}(\|\Sigma_A - \nu\mathbb{I}_A\|_1 \geq \nu\epsilon'(\epsilon, k)). \tag{87}$$

This concludes the proof of Lemma 4. ∎

*Proof of Theorem 2.* Finally, by utilizing Lemma 3 and Lemma 4, we can prove Theorem 2,

$$\begin{aligned}
\mathbb{P}(D_{\mathrm{KL}}(\mathcal{N}_A||\mathcal{N}_\nu) \geq \epsilon) &\leq \mathbb{P}(\|\Sigma_A - \nu\mathbb{I}_A\|_1 \geq \nu\epsilon'(\epsilon, k)) \\
&\leq C\frac{1+\frac{\nu\epsilon'}{2k}}{\nu^2\epsilon'^2 n} + De^{-n\nu^2\epsilon'^2 F}.
\end{aligned} \tag{88}$$

Here $C$ is the same constant in Theorem 1, $D = 8k^2$, and $F = 1/(1024e^{4s}k^2)$. □

## V. DETAILS OF NUMERICS IN BRICKWORK LINEAR-OPTICAL CIRCUIT MODEL

In this section, we provide details on the numerics involving the brickwork linear-optical circuit model which gave rise to Fig. 3 of the main text. We also provide additional numeric simulations supporting the universality of deep thermalization to the Gaussian Scrooge distribution in Gaussian continuous-variable systems, as well as investigations into when it breaks down.

### A. Linear-optical circuit model

As described in the main text, the linear-optical circuit model consists of a 1d array of $n$ bosonic modes, initially prepared in a product squeezed state with uniform squeezing $s$ (we always pick $s = 1/2$ in what follows), and coupled via repeated applications of beam-splitters and phase-shifters. Specifically, we consider a brickwork architecture wherein odd pairs of nearest-neighbor modes first couple via two-mode beam-splitters, followed by a single-mode phase-shift on the left mode (denoted by the ortho-symplectic operators $\mathrm{BS}_{i,i+1}$ and $\mathrm{PS}_i^\theta = \begin{pmatrix} \cos\theta_i & -\sin\theta_i \\ \sin\theta_i & \cos\theta_i \end{pmatrix}$ respectively in phase space; here $\theta_i$ is the value of the phase shift). Then, even pairs of nearest-neighbor modes couple in an identical fashion. This defines one 'layer' of the circuit. The overall circuit then consists of $t$ applications of such layers, which we interpret as 'time'. We will consider different scenarios of uniform beam-splitting and phase-shifts within each layer or not (i.e., spatial uniformity/randomness), across layers (i.e., temporal uniformity/randomness), or a mixture of both (i.e., spatio-temporal uniformity/randomness). We will also consider different boundary conditions (open vs periodic). The projected ensemble is constructed from the middle $k$ modes of the chain, assuming Gaussian measurements on the complementary subsystem parameterized by squeezing parameter $s_\sigma$. In our numerics, we always pick $k = 3$.

### B. Open boundary conditions, 50:50 beam-splitter, uniform phase-shift in space and time

To begin, we consider the simplest scenario of circuit evolution with equal-weight (i.e., 50:50) beam-splitters

$$\mathrm{BS}_{i,i+1} = \frac{1}{\sqrt{2}}\begin{pmatrix} 1 & 1 & 0 & 0 \\ -1 & 1 & 0 & 0 \\ 0 & 0 & 1 & 1 \\ 0 & 0 & -1 & 1 \end{pmatrix}_{i,i+1}, \tag{89}$$

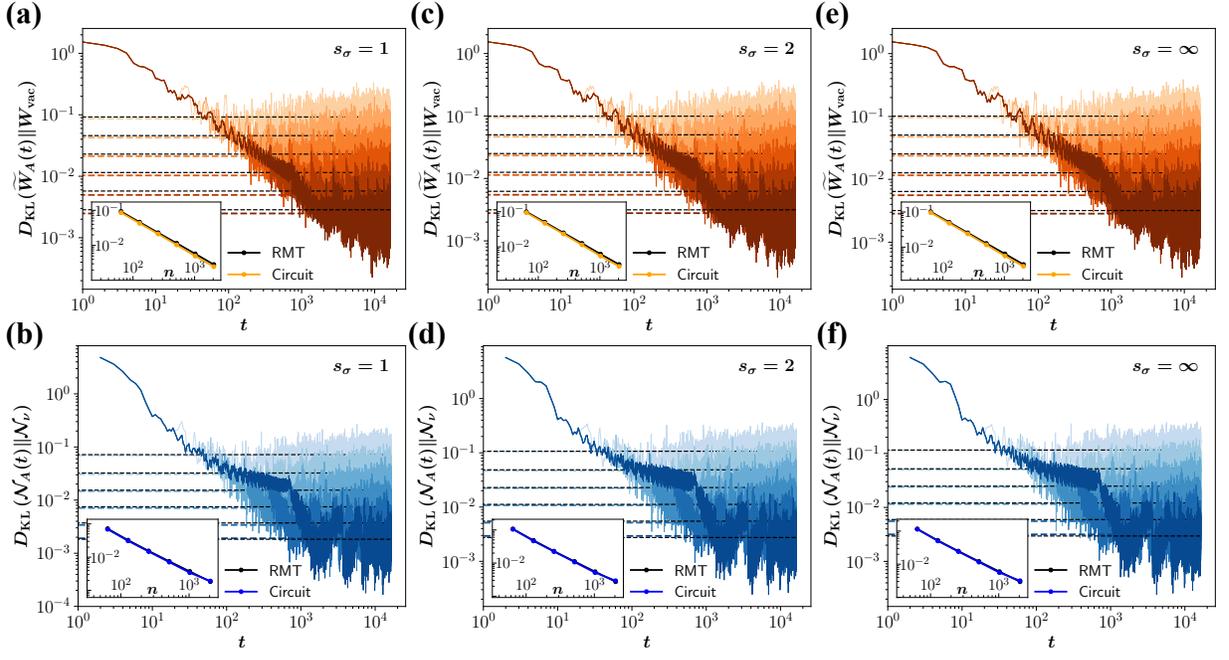

Figure S1. KL-divergences over time of (a,c,e) the Wigner functions of a 3-mode projected state $W_A(t)$ from a coherent state of equal displacement $W_{\text{vac}}$, and (b,d,f) distribution of displacements $\mathcal{N}_A(t)$ from the expected universal distribution $\mathcal{N}_\nu$, with generalized heterodyne measurement bases characterized by $s_\sigma = 1, 2, \infty$, respectively. Here the brickwork-circuit model is defined with open boundary conditions, equal-weight (50:50) beam-splitters, and uniform phase shift $\theta = \pi/8$ across space and time. Lighter to darker shades represent data for different system sizes $n = 64, 128, 256, 512, 1024, 2048$. Insets show comparison of the saturated values to those generated from corresponding globally random bosonic Gaussian states; the agreements are excellent.

(expressed under the basis $(\hat{q}_i, \hat{q}_{i+1}, \hat{p}_i, \hat{p}_{i+1})$) and phase shift $\theta_i = \theta = \pi/8$, uniform across both space and time. These are the parameters used to generate Fig. 3 of the main text, where the measurement basis is that of unsqueezed coherent states $s_\sigma = 0$. There we found that our claim of the limiting form of the PE to the Gaussian Scrooge distribution $\mathcal{N}_A(t) \to \mathcal{N}_\nu$, $\widetilde{V}_A(t) \to \mathbb{I}_A$ arises in the large $n$ and large $t$ limit (taken in that order). Specifically, we saw that for any fixed $n$, the KL divergences of either the Wigner function of the projected state from an unsqueezed coherent state of equal displacement, or the distribution of displacements $\mathcal{N}_A$ from the expected $\mathcal{N}_\nu$, decay as $\sim 1/t$, before saturating to a value scaling as $1/n$. Moreover, we find that this saturation value agrees very well with that of the average KL divergences from the PE generated from an $n$-mode random BGS. In Fig. S1, we repeat the same analysis for other measurement bases $s_\sigma = 1, 2, \infty$. Again, we observe similar behavior as in Fig. 3 of the main text, showing the independence of the phenomena to measurement bases and hence supporting the universality of our claim. We also note that this behavior is insensitive to the value of the phase shift (modulo special values like $\theta = m\pi/4, m \in \mathbb{Z}$ where the system is integrable).

### C. Open boundary conditions, 30:70 beam-splitter, uniform phase-shift in space and time

To further corroborate the universality of our claim, we investigate the sensitivity of the limiting ensemble of the PE to different choices of beam-splitters. In Fig. S2, we repeat the same analysis as in Sec. V B, but assume a biased 30:70 beam-splitter defined by

$$\text{BS}_{i,i+1} = \begin{pmatrix} \cos\phi & \sin\phi & 0 & 0 \\ -\sin\phi & \cos\phi & 0 & 0 \\ 0 & 0 & \cos\phi & \sin\phi \\ 0 & 0 & -\sin\phi & \cos\phi \end{pmatrix}, \tag{90}$$

where $\phi = \arccos(\sqrt{0.3})$. Again, we see similar behavior of convergence to the limiting behavior of the Gaussian Scrooge distribution as in Fig. S1.

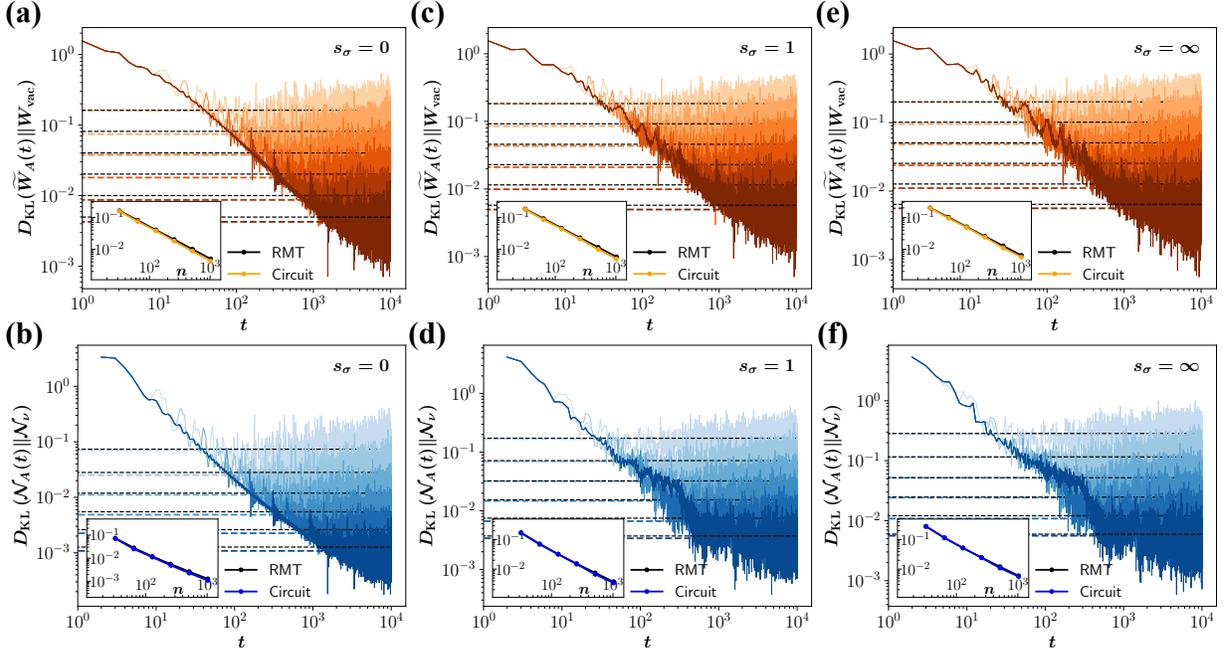

Figure S2. KL-divergences over time of (a,c,e) the Wigner functions of a 3-mode projected state $W_A(t)$ from a coherent state of equal displacement $W_{\text{vac}}$, and (b,d,f) distribution of displacements $\mathcal{N}_A(t)$ from the expected universal distribution $\mathcal{N}_\nu$, with generalized heterodyne measurement bases characterized by $s_\sigma = 0, 1, \infty$, respectively. Here the brickwork-circuit model is defined with open boundary conditions, biased (30:70) beam-splitters, and uniform phase shift $\theta = \pi/8$ across space and time. Lighter to darker shades represent data for different system sizes $n = 32, 64, 128, 256, 512, 1024$. Insets show comparison of the saturated values to those generated from corresponding globally random bosonic Gaussian states; the agreements are excellent.

## D. Periodic boundary conditions, 50:50 beam-splitter, uniform phase-shift in space and time

We also investigate the role of boundary conditions in determining the convergence of the PE to the expected limiting form of the Gaussian Scrooge distribution. Here we repeat the same simulations as in Sec. V B but assume periodic boundary conditions. Fig. S3 shows the results. Intriguingly, while for measurement basis $s_\sigma = 0$ we again see convergence, for measurement bases $s_\sigma \neq 0$ we observe deviations. In particular, we note that the KL divergences (of either the Wigner function or the distribution of displacements) are not decaying to zero for large $n$ and large $t$, but rather appear to be saturating to a non-zero value. This suggests that the limiting form of the PE is *not* the Gaussian Scrooge distribution in those cases.

What could be the physical reason for the observed deviations? Let us first note that the reduced density matrix $\hat{\rho}_A$ is identical in all cases (as it is the state of the subsystem *ignorant* of the measurement outcomes), and *is* characterized by the convergence of its covariance matrix to $V_A \to (2\nu+1)\mathbb{I}_A$ (we numerically find this to be true always). Thus, the lack of convergence of the projected ensemble to the Gaussian Scrooge distribution cannot be explained by the system not thermalizing in the 'regular' sense; instead, it is the failure to deep thermalize. Note that one of the special properties of the Gaussian Scrooge distribution is that measurement outcomes $\mathbf{r}_B$ on the bath $B$ are minimally correlated with the projected state $|\psi(\mathbf{r}_A(\mathbf{r}_B), \tilde{V}_A)\rangle$ (a consequence of the quantum information-theoretic property of having *minimal accessible information*, see Section VI). The lack of emergence of the GSD suggests that in the case of the circuit with periodic boundary conditions, the measurements on $B$ in squeezed bases $s_\sigma \neq 0$ are somehow *learning* about the state of $A$. Now, the only difference between the OBC case of Sec. V B and the PBC case of this section, is in the translational symmetry (or lack thereof) of the circuit: in the latter, we expect that eigenstates of the circuit can be further labeled by quasimomenta, while in the former they cannot. We thus hypothesize that the reason for the lack of convergence to the Gaussian Scrooge distribution observed here is that measurements in the squeezed-state bases $s_\sigma \neq 0$ have non-trivial overlap with the momentum modes of the system, much like how measurements in spin-systems with energy conservation, in a basis which is correlated with the energy operator, will not produce a projected ensemble that tends to the (spin) Scrooge distribution but rather the 'generalized Scrooge distribution' [11]. We leave further investigation of this interesting generalized-deep-thermalizing scenario to future work.

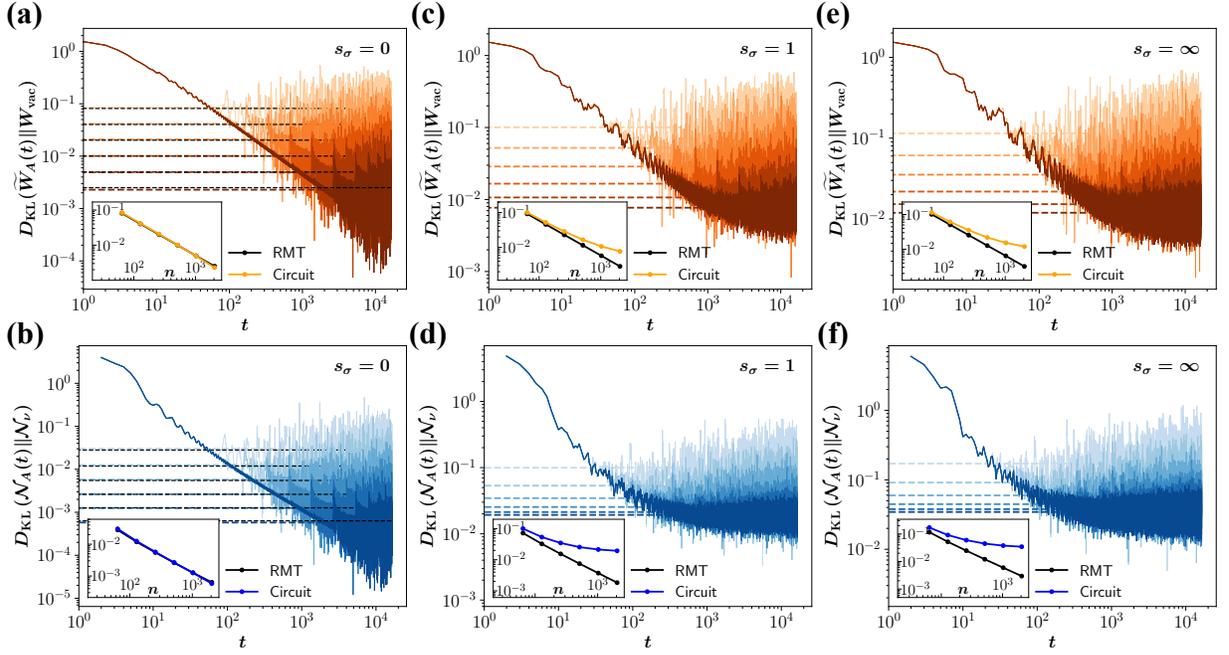

Figure S3. KL-divergences over time of (a,c,e) the Wigner functions of a 3-mode projected state $W_A(t)$ from a coherent state of equal displacement $W_{\text{vac}}$, and (b,d,f) distribution of displacements $\mathcal{N}_A(t)$ from the expected universal distribution $\mathcal{N}_\nu$, with generalized heterodyne measurement bases characterized by $s_\sigma = 0, 1, \infty$, respectively. Here the brickwork-circuit model is defined with periodic boundary conditions, equal-weight (50:50) beam-splitters, and uniform phase shift $\theta = \pi/8$ across space and time. Lighter to darker shades represent data for different system sizes $n = 64, 128, 256, 512, 1024, 2048$. For general squeezed measurements $s_\sigma \neq 0$, the system apparently fails to deep thermalize, even though it thermalizes 'regularly'. Insets show comparison of the saturated values to those generated from corresponding globally random bosonic Gaussian states.

### E. 50:50 beam-splitter, phase-shift random in space but uniform in time

Finally, we study a circuit with 50:50 beam-splitters, but assume the phase-shifts are random ($\theta_i \sim \text{Uniform}[0, 2\pi]$) across space while fixed over time. We find boundary conditions do not matter in this scenario. Fig. S4 shows the results. We observe that the system does not thermalize in the regular sense ($V_A \not\to (2\nu + 1)\mathbb{I}_A$), let alone deep thermalize. We attribute this to *localization* of the eigenmodes of the 1-layer evolution operator (of the circuit) due to the disorder. Interestingly, we find that localization as a failure mode for the system to thermalize (or deep-thermalize) is sensitive to the fact that the circuit is comprised of *identical* layers repeated in time, i.e., has *discrete-time periodicity*. In such a case, the system can be viewed as 'Floquet', and has a conserved quasi-energy operator due to Floquet's theorem. If we instead break this time-periodicity (such as with temporal randomness; or by using a deterministic but aperiodic drive, like a circuit wherein different layers correspond to one of two fixed but distinct evolution operators, alternating according to the characters of the Fibonacci word [12]), then convergence to the Gaussian Scrooge distribution will be seen again (data not shown). This is despite the fact that within each layer there can still be spatial randomness in the phase shifts resulting in localization. That is, the presence of localization *and* quasi-energy conservation is necessary for regular thermalization (and deep thermalization) to be arrested; while just localization (of the single-layer evolution operator at every time-step) is not sufficient (indeed note that Ref. [12] rigorously showed that the Fibonacci drive, despite being quasiperiodic, does not admit a conserved quasi-energy operator).

## VI. MAXIMUM ENTROPY PRINCIPLE AND GAUSSIAN SCROOGE DISTRIBUTION

In this section, we provide a detailed discussion of our proposed maximum entropy principle, which is used to predict the limiting form of a quantum state ensemble arising from suitably complex interactions/dynamics. We will first restate the principle for general physical systems, recap what it predicts for finite-dimensional (i.e., spin) quantum systems — namely, the emergence of the so-called *Scrooge distribution*, before analyzing the case of Gaussian continuous-variable quantum systems, where we show that the maximum-entropy ensemble is the so-called *Gaussian Scrooge distribution* (Theorem 3).

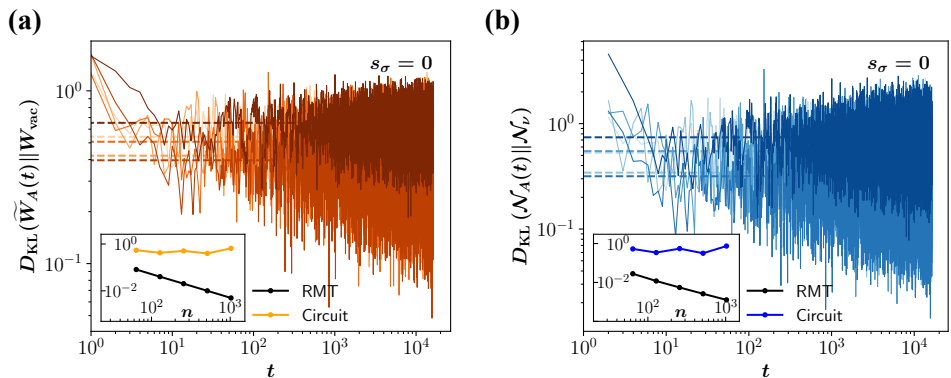

Figure S4. KL-divergences over time of (a) the Wigner functions of a 3-mode projected state $W_A(t)$ from a coherent state of equal displacement $W_{vac}$, and (b) distribution of displacements $\mathcal{N}_A(t)$ from the expected universal distribution $\mathcal{N}_\nu$, with coherent-state measurement bases $s_\sigma = 0$. Here the brickwork-circuit model is defined with periodic boundary conditions, while phase shifts $\theta$ are uniformly random in space but constant over time. Lighter to darker shades represent data for different system sizes $n = 64, 128, 256, 512, 1024$. In this scenario, the system does not regularly thermalize, let alone deep thermalize. Insets show a comparison of the saturated values to those generated from corresponding globally random bosonic Gaussian states.

### A. Maximum entropy principles in thermalization and deep thermalization

Maximum entropy principles correspond to general powerful propositions allowing one to predict the state of a system with incomplete data: they assert that the probability distribution that best describes a system should be the one that agrees with given available knowledge, but assumes the least about other properties, quantitatively captured by the maximization of some *information entropy*. Such principles appear in many different contexts, like information theory and statistical thermodynamics. A key aspect governing the success of maximum entropy principles is in identifying and justifying the appropriate information entropy to use in different physical scenarios.

In the context of quantum statistical physics — the arena of this work, the physical question that naturally arises where a maximum entropy principle is useful is this: given a global quantum state on many constituents, arising from some complex set of interactions (like eigenstates of a quantum many-body Hamiltonian or states evolved for long times under quantum chaotic dynamics), what is the *local* state of a system? In other words, what form does the reduced density matrix (RDM) $\hat{\rho}_A$ on a local subsystem $A$ assume? This is the question of quantum thermalization. The information entropy which has been understood to be relevant in this scenario, is that of the von Neumann entropy $S = -\text{Tr}(\hat{\rho}_A \log \hat{\rho}_A)$, since this is relevant to the thermodynamic free energy of subsystem $A$ [13]. Then, the maximum entropy principle for regular quantum thermalization states: the subsystem $A$ should be described by a density matrix which maximizes the von Neumann entropy $S$ subject to any conservation laws (like conservation of charge or energy). Predictions about the equilibrium expectation values of any local observables then follow. For example, it is a standard textbook exercise to derive that the thermal Gibbs state $\hat{\rho}_{th} \propto e^{-\beta \hat{H}}$ at inverse temperature $\beta$ defined as $E = -\partial_\beta \log \text{Tr}(e^{-\beta \hat{H}})$, is the unique density matrix maximizing the entropy over all density matrices $\hat{\rho}$ with the same mean energy $E = \text{Tr}(\hat{\rho}\hat{H})$. Note importantly that the maximum entropy principle is but an appealing *guiding principle* and not derived from first principles, so may not hold in all cases. As it is believed that *ergodicity* of dynamics is what microscopically underpins the success of the principle, exceptions to the predictions of the maximum entropy principle are typically attributed to a breaking of ergodicity due to some physical mechanism (such as localization, quantum many-body scarring, or Hilbert space fragmentation/shattering).

Turning now to the projected ensemble (PE) $\mathcal{E}$, which is a collection of pure quantum states of a local subsystem each of which represents the subsystem's state *conditioned* on a measurement outcome on the complement, the physical question is: what should the limiting form of the PE be under complex interactions or dynamics? This is the question of *deep thermalization*. Note that even though the PE is technically constructed by associating each pure state $|\psi\rangle$ on $A$ with a measurement outcome on $B$, it is mathematically equivalent to consider the PE simply as a distribution solely on $A$,

$$\mathcal{E} = \{p_\mathcal{E}(\psi), |\psi\rangle\}, \tag{91}$$

where $p_\mathcal{E}(\psi)$ is underlying probability distribution on the states (formally, we construct $p_\mathcal{E}(\psi)$ via $p_\mathcal{E}(\psi) = \sum_z p(z)\delta(|\psi\rangle\langle\psi| - |\psi_z\rangle\langle\psi_z|)$ where $p(z)$ is the Born probability of measurement outcome $z$ and $|\psi_z\rangle$ the corresponding conditional state). We wish to formulate a maximum entropy principle for the projected ensemble too, but two questions immediately arise: (i) what is the appropriate information entropy to use? (ii) what is knowledge that

can be assumed already known, which constrains the optimization problem?

To answer the first question, we argue that information of the pure state ensemble Eq. (91), a classical-quantum object, is contained naturally in two places: (a) the distribution $p_{\mathcal{E}}(\psi)$ of states over the Hilbert space, and also (b) information extractable from a quantum measurement on the conditional quantum states itself. That is to say, we can define a *joint probability distribution* $p_{\mathcal{E},\mathcal{M}}$ of the ensemble and some choice of measurement (a POVM) $\mathcal{M} = \{\hat{M}_j\}$

$$p_{\mathcal{E},\mathcal{M}}(\psi_i, \hat{M}_j), \tag{92}$$

from which we can also define the marginal distributions $p_{\mathcal{E}}(\psi_i)$ and

$$p_{\mathcal{M}}(\hat{M}_j) = \text{Tr}(\hat{\rho}_{\mathcal{E}} \hat{M}_j), \tag{93}$$

where $\hat{\rho}_{\mathcal{E}} = \sum_i p_{\mathcal{E}}(\psi_i)|\psi_i\rangle\langle\psi_i|$ is the first statistical moment of the state ensemble, i.e., its density matrix. (Note the indices $i, j$ can take discrete or continuous values; here we temporarily switch to a discrete notation for convenience). Now, we appeal to the physical intuition that if interactions or dynamics are complex enough, measurements on a conditional state $|\psi\rangle$ should reveal very little about which measurement on the bath (the complementary subsystem where measurements are performed) produced it in the first place, due to information scrambling; in other words, the state ensemble $\mathcal{E}$ and measurement apparatus $\mathcal{M}$ should be least correlated. This motivates us to define the *ensemble entropy* that should be maximized in a maximum entropy principle to be

$$\mathcal{S}(\mathcal{E}) := - \sup_{\mathcal{M} \in \text{POVM}} D_{\text{KL}}(p_{\mathcal{E},\mathcal{M}} \| p_{\mathcal{E}} \otimes p_{\mathcal{M}}), \tag{94}$$

where $\mathcal{M}$ runs over all POVMs. Here $D_{\text{KL}}(p_{\mathcal{E},\mathcal{M}} \| p_{\mathcal{E}} \otimes p_{\mathcal{M}})$ refers to the Kullback-Leibler divergence, or *relative entropy* of the joint distribution $p_{\mathcal{E},\mathcal{M}}$ from the product distribution $p_{\mathcal{E}} \otimes p_{\mathcal{M}} = p_{\mathcal{E}} p_{\mathcal{M}}$. We note that $-D_{\text{KL}}(p_{\mathcal{E},\mathcal{M}} \| p_{\mathcal{E}} \otimes p_{\mathcal{M}})$ can be interpreted as the number of classical bits needed to encode the joint distribution of $p_{\mathcal{E},\mathcal{M}}$, according to an optimal scheme assuming prior knowledge only of the marginals (more precisely, after adding an appropriate positive constant, see [11]). The supremum over POVMs has the effect of ensuring the entropy so-defined is a property solely of the state ensemble, wherein we choose the optimal POVM to give the best distinguishability. Note that $S(\mathcal{E}) \leq 0$ and is zero if and only if $p_{\mathcal{E},\mathcal{M}} = p_{\mathcal{E}} p_{\mathcal{M}}$.

We turn next to the question of how much knowledge should be deemed available in the application of the maximum entropy principle for deep thermalization. We note that under similar considerations of complex interactions/dynamics, the maximum entropy principle for regular quantum thermalization already fixes the density matrix $\hat{\rho}_A$ of the subsystem (see above discussion; for example, it should be the thermal Gibbs state in the presence of energy conservation), so it is natural to take this information as a constraint in the optimization.

To recapitulate, we therefore propose that the maximum entropy principle for deep thermalization is as such: *the limiting projected ensemble $\mathcal{E}$ obtained under complex interactions should have maximal ensemble entropy $\mathcal{S}(\mathcal{E})$, subject to fixed first moment $\hat{\rho}_A$, determined by regular thermalization.*

Before deriving the consequences of this principle, we note that there is an intimate connection between the *ensemble entropy* $S(\mathcal{E})$ and a quantum information-theoretic property of the ensemble of states called the *accessible information* $I(\mathcal{E})$. Indeed, $-S(\mathcal{E})$ is nothing more than the ensemble's accessible information, defined as

$$I(\mathcal{E}) := \sup_{\mathcal{M} \in \text{POVM}} I(\mathcal{E} : \mathcal{M}), \tag{95}$$

where $I(\mathcal{E} : \mathcal{M})$ is the mutual information

$$I(\mathcal{E} : \mathcal{M}) := H(\psi) + H(M) - H(\psi, M) \tag{96}$$
$$= D_{\text{KL}}(p_{\mathcal{E},\mathcal{M}} \| p_{\mathcal{E}} p_{\mathcal{M}}) \tag{97}$$

and $H(\cdot)$ is the Shannon entropy. The accessible information of the ensemble $\mathcal{E}$ has the operational meaning of the maximum amount of classical information extractable from quantum measurements on quantum states of $\mathcal{E}$, when they are used to encode a message through a quantum channel. Thus, we may gain an alternative quantum-information perspective to the proposed maximum entropy principle of deep thermalization: the bath $B$ is encoding information about its particular state within the conditional states $|\psi\rangle$ and sending it to $A$, but it does so in a way in which information is maximally difficult to extract from the local subsystem, in line with the intuition that complex quantum dynamics scrambles information and hides it very well from local observers.

### B. Scrooge distribution maximizes the ensemble entropy in finite-dimensional (spin) systems

As a warm-up, we first apply the maximum entropy principle for state-ensembles in *finite-dimensional* (i.e., spin) systems, to determine the universal form of the projected ensemble that emerges. It turns out that the question of the pure state ensemble $\mathcal{E}$ whose first moment is $\hat{\rho}$ and which minimizes its accessible information (equivalently, maximizes its ensemble entropy) was solved already by Jozsa et al. [14] for spin systems. They showed that a quantity known as the subentropy $Q(\hat{\rho})$ (a function only of $\hat{\rho}$) lower bounds the accessible information $I(\mathcal{E})$ for *any* ensemble $\mathcal{E}$ under the constraint that they have the same first moment $\hat{\rho}_{\mathcal{E}} = \hat{\rho}$,

$$Q(\hat{\rho}) \leq I(\mathcal{E}), \tag{98}$$

and they also showed that this lower bound is attained within the *Scrooge distribution* (so-called because it is most 'stingy' with its information):

$$\mathcal{E}_{\text{Scr.}} = \left\{ \mathrm{d}\phi D \langle \phi | \hat{\rho} | \phi \rangle, \frac{\sqrt{\hat{\rho}} | \phi \rangle}{\sqrt{\langle \phi | \hat{\rho} | \phi \rangle}} \right\}, \tag{99}$$

where $\mathrm{d}\phi$ is the uniform Haar measure on the Hilbert space and $D$ the dimension. Note the Scrooge distribution has a number of equivalent forms, one of which is the *Gaussian adjusted projected* (GAP) measure [14–16]. From the expression, we may understand the Scrooge distribution as a '$\hat{\rho}$-distortion' of the uniform Haar distribution (for example, set $\hat{\rho} = \hat{I}/D$ as the limiting case to see this).

In the context of complex dynamics, since we expect $\hat{\rho}$ to be a thermal Gibbs state arising from regular thermalization, the limiting ensemble in deep thermalization should thus be the Scrooge ensemble wherein its first moment is the thermal density matrix. Indeed, previous studies [11, 17–19] have corroborated the emergence of such a distribution, in line with the predictions of the proposed maximum entropy principle. We stress though that just like in the case of the maximum entropy principle of regular thermalization, the maximum entropy principle for state ensembles cannot be expected to hold in all cases, and elucidating when and why it fails is an interesting direction of exploration (for example, [11] showed that if the measurement basis on $B$ has a large overlap with eigenstates of conserved quantities of the system, such that the hypothesis of the system and bath becoming uncorrelated should not be expected to hold, then a generalized version of the Scrooge distribution instead appears).

### C. Gaussian Scrooge distribution maximizes the ensemble entropy in Gaussian CV systems

We now return to the focus of this work: the form of the deep thermalized projected ensemble in continuous-variable (CV) quantum systems, assuming Gaussian states and Gaussian measurements. We wish to employ our aforementioned maximum entropy principle to predict the limiting state ensembles in this setting. However, Jozsa et al.'s construction of the Scrooge ensemble does not generalize, as there is no obvious notion of a *uniform* (normalized) distribution of bosonic Gaussian states in Hilbert space, owing to the unboundedness of the Hilbert space.

Nevertheless, building off recent seminal works of Holevo [20, 21] bounding and computing accessible information of Gaussian state ensembles, we are able to ascertain the Gaussian pure state ensemble which minimizes the accessible information (equivalently, maximizes the ensemble entropy), in the case that the reduced density matrix is a thermal Gaussian state (i.e., its covariance matrix is proportional to $\mathbb{I}$), which we call in analogy the *Gaussian Scrooge distribution*. Concretely, we have:

**Theorem 3.** *Consider the set of $k$-mode bosonic Gaussian pure state ensembles of the form $\mathcal{E}_{\Sigma, \widetilde{V}} = \{p_\Sigma(\mathbf{r}) \mathrm{d}\mathbf{r}, |\psi(\mathbf{r}, \widetilde{V})\rangle\}$ such that each ensemble's density matrix $\hat{\rho}$ is equal, characterized by having zero displacement and covariance $2\Sigma + \widetilde{V} = (2\nu + 1)\mathbb{I}$ for some $\nu \geq 0$. Here $p_\Sigma(\mathbf{r})$ denotes the probability density of a centered multivariate Gaussian distribution with covariance matrix $\Sigma$ over the phase space, and $|\psi(\mathbf{r}, \widetilde{V})\rangle$ the bosonic Gaussian pure state with random displacement $\mathbf{r}$ and common covariance matrix $\widetilde{V}$. Then the ensemble which maximizes [22] the ensemble entropy $S(\mathcal{E}_{\Sigma, \widetilde{V}})$ is given by*

$$\Sigma = \nu \mathbb{I}, \quad \widetilde{V} = \mathbb{I}, \tag{100}$$

*which we call the Gaussian Scrooge distribution (GSD), and the maximum entropy is*

$$S_{\max}(\mathcal{E}_{\Sigma, \widetilde{V}}) = -k \log(1 + \nu). \tag{101}$$

*Equivalently, the GSD can be written as a '$\hat{\rho}$-distortion' of unsqueezed coherent states*

$$\mathcal{E}_{\text{GSD}} = \left\{ \mathrm{d}\mathbf{r}\, p(\mathbf{r}), \sqrt{\hat{\rho}}|\mathbf{r}\rangle / \sqrt{\langle\mathbf{r}|\hat{\rho}|\mathbf{r}\rangle} \right\}, \tag{102}$$

*where $p(\mathbf{r}) = \langle\mathbf{r}|\hat{\rho}|\mathbf{r}\rangle/(2\pi)^k$ and $|\mathbf{r}\rangle = |\psi(\mathbf{r},\mathbb{I})\rangle$ is an unsqueezed coherent state on with displacement $\mathbf{r} \in \mathbb{R}^{2k}$.*

Here, we have framed the theorem directly in terms of the expected thermal Gaussian density matrix $\hat{\rho} = \hat{\rho}_{\text{th}}$ arising from regular thermalization, which has covariance $2\Sigma + \widetilde{V} = (2\nu+1)\mathbb{I}$ ($\nu$ being the particle-number density). Note that the set of Gaussian ensembles $\mathcal{E}_{\Sigma,\widetilde{V}}$ we optimize over, reflects the best of our knowledge of what its form should be while being completely ignorant of what the precise global generator state: we only know that (i) the particle-number is conserved, captured by $\nu$; (ii) the projected states $|\psi(\mathbf{r},\widetilde{V})\rangle$ are *pure* due to the choice of rank-1 measurements, captured by the covariance matrix $\widetilde{V}$ being both positive and symplectic; and (iii) the covariance matrix of the projected states are *common*, i.e. independent of displacement $\mathbf{r}$ (which happens to follow a multivariate normal distribution), easily seen from its defining expression from the Born rule. To solve the constrained optimization problem, we exploit the following result of Holevo's, which explicitly computes the accessible information for a Gaussian state ensemble, not necessarily pure:

**Lemma 5. (Theorem 2, Holevo [21]).** *For a $k$-mode Gaussian state ensemble $\mathcal{E}_{\Sigma,\widetilde{V}} = \{p_\Sigma(\mathbf{r})\mathrm{d}\mathbf{r}, \hat{\rho}(\mathbf{r},\widetilde{V})\}$, where $p_\Sigma(\mathbf{r})$ is a centered multivariate normal distribution with covariance $\Sigma$ and $\hat{\rho}(\mathbf{r},\widetilde{V})$ are Gaussian states (not necessarily pure) with displacement $\mathbf{r}$ and common covariance $\widetilde{V}$, its accessible information is equal to*

$$I(\mathcal{E}_{\Sigma,\widetilde{V}}) = \frac{1}{2}\log\det\left(V + \Xi\right)\left(\Xi + \Omega J_\Xi\right)^{-1}, \tag{103}$$

*where*

$$V = \widetilde{V} + 2\Sigma, \quad \Xi = \frac{1}{2}V\Upsilon^T\Sigma^{-1}\Upsilon V - V, \quad \Upsilon = \sqrt{\mathbb{I}_{2k} + (V\Omega^{-1})^{-2}}, \tag{104}$$

*given that the threshold condition*

$$V - \Omega J_\Xi \geq 0 \tag{105}$$

*holds. Here, $J_\Xi$ denotes the complex structure of $\Xi$.*

The complex structure of a positive operator is discussed in Section VI E, and is related to the Williamson decomposition (also known as symplectic diagonalization) of a positive matrix. If $\Xi$ is the covariance matrix of a Gaussian state (which turns out to be true for our case of interest), the Heisenberg uncertainty relation can be equivalently written as $\Xi - \Omega J_\Xi \geq 0$, with the inequality saturated (i.e., $\Xi = \Omega J_\Xi$) if and only if $\Xi$ is the covariance matrix of a Gaussian pure state. We will make use of this fact in the following derivations.

*Proof of Theorem 3.* Consider the special case of Gaussian *pure* state ensembles $\mathcal{E}_{\Sigma,\widetilde{V}} = \{p_\Sigma(\mathbf{r})\mathrm{d}\mathbf{r}, |\psi(\mathbf{r},\widetilde{V})\rangle\}$ in Lemma 5. The constraint of quantum thermalization in the premise of Theorem 3 reads $V = \widetilde{V} + 2\Sigma = (1+2\nu)\mathbb{I}$ which implies

$$\Upsilon = \sqrt{1 - \frac{1}{(1+2\nu)^2}}\mathbb{I}, \tag{106}$$

and

$$\Xi = 2\nu(1+\nu)\Sigma^{-1} - (1+2\nu) = \frac{-1 + (1+2\nu)\widetilde{V}}{1+2\nu-\widetilde{V}}. \tag{107}$$

We know both $\Sigma$ and $\widetilde{V}$ are positive, so $\widetilde{V} \leq (1+2\nu)\mathbb{I}$. We claim that $\Xi$ is the covariance matrix of a Gaussian pure state. This can be argued in two steps:

- Since $\widetilde{V}$ represents a pure state (positive and symplectic) with constraints, by the Bloch-Messiah decomposition, $\widetilde{V}$ can be diagonalized as $\widetilde{V} = O[D \oplus D^{-1}]O^T$ where $D = \text{diag}(x_1, x_2, \cdots, x_k)$ with $1 \leq x_i \leq (2\nu+1)$ for each $i$, and $O$ is an ortho-symplectic matrix. It follows that the eigenvalues of $\widetilde{V}$ come in pairs $\{x_i, x_i^{-1}\}_{i=1}^k$.

- By Eq. (107), we know that $\Xi$ and $\widetilde{V}$ are simultaneously diagonalizable. We may thus directly calculate $\Xi = O\left(\mathrm{diag}\left[\xi(x_1), \cdots, \xi(x_k), \xi(x_1^{-1}), \cdots, \xi(x_k^{-1})\right]\right) O^T$ where

$$\xi(x) \equiv \frac{-1 + (1+2\nu)x}{1 + 2\nu - x}. \tag{108}$$

It can then be easily checked that for any $x \geq 1$, $\xi(x) \geq 1$, and that $\xi(x) \cdot \xi(x^{-1}) = 1$.

Therefore, $\Xi$ is both symplectic and positive, and could represent a pure Gaussian state. Thus $\Omega J_{\Xi} = \Xi$. But the threshold condition Eq. (105), which in this case can be written as $(1+2\nu)\mathbb{I} \geq \Xi$, is not necessarily true since $\xi(x)$ is unbounded on $x \in [1, 1+2\nu)$. By substituting $\Omega J_{\Xi}$ with $\Xi$, the accessible information of $\mathcal{E}_{\Sigma, \widetilde{V}}$ is

$$
\begin{aligned}
I(\mathcal{E}_{\Sigma, \widetilde{V}}) &= \frac{1}{2} \log \det\left(\frac{V + \Xi}{2\Xi}\right) = \frac{1}{2} \log \det\left(\frac{2\nu(1+\nu)}{(1+2\nu)\widetilde{V} - 1}\right) \\
&= \frac{1}{2} \log\left(\prod_{i=1}^{k} \frac{2\nu(1+\nu)}{(1+2\nu)x_i - 1} \frac{2\nu(1+\nu)}{(1+2\nu)\frac{1}{x_i} - 1}\right) \\
&= \frac{1}{2} \log\left(\prod_{i=1}^{k} \frac{4\nu^2(1+\nu)^2}{(1+2\nu)^2 - (1+2\nu)(x_i + \frac{1}{x_i}) + 1}\right) \geq k \log(1+\nu),
\end{aligned}
\tag{109}
$$

where the inequality in the last line is saturated if and only if $x_i = 1$ for all $i$ (i.e., $\widetilde{V} = \mathbb{I}$, and consequently, $\Sigma = \nu\mathbb{I}$). This is equivalently $S(\mathcal{E}_{\Sigma, \widetilde{V}}) \equiv -I(\mathcal{E}_{\Sigma, \widetilde{V}}) \leq -k \log(1+\nu)$. Furthermore, one can easily check that the threshold condition holds in the vicinity of the optimal solution ($\Sigma = \nu\mathbb{I}, \widetilde{V} = \mathbb{I}$).

Finally, we prove the last statement of the theorem, that the GSD can be written as

$$\mathcal{E}_{\mathrm{GSD}} = \left\{ \mathrm{d}\mathbf{r} p(\mathbf{r}), \sqrt{\hat{\rho}}|\mathbf{r}\rangle / \sqrt{\langle \mathbf{r}|\hat{\rho}|\mathbf{r}\rangle} \right\} \tag{110}$$

where $\hat{\rho} = \hat{\rho}_{\mathrm{th}} \propto e^{-\beta \hat{N}}$ is the thermal Gaussian state with $\nu^{-1} = (e^{\beta} - 1)$ such that it has zero displacement and covariance $(2\nu + 1)\mathbb{I}$, while $|\mathbf{r}\rangle$ is an unsqueezed coherent state with displacement $\mathbf{r} \in \mathbb{R}^{2k}$. To see this, we note that the $\hat{\rho}_{\mathrm{th}}$-distortion of a coherent state $|\mathbf{r}\rangle$ remains an unsqueezed coherent state:

$$\sqrt{\hat{\rho}_{\mathrm{th}}}|\mathbf{r}\rangle \propto |\mathbf{r}e^{-\beta/2}\rangle. \tag{111}$$

We can then calculate

$$p(\mathbf{r}) \propto \exp\left(-\frac{1}{2(\nu+1)}\mathbf{r}^T\mathbf{r}\right). \tag{112}$$

By doing a simple change of variables we can equivalently write

$$\mathcal{E}_{\mathrm{GSD}} = \{\mathrm{d}\mathbf{r} p'(\mathbf{r}), |\mathbf{r}\rangle\} \tag{113}$$

where

$$p'(\mathbf{r}) \propto \exp\left(-\frac{1}{2\nu}\mathbf{r}^T\mathbf{r}\right), \tag{114}$$

and so we see the GSD is composed of unsqueezed coherent states with displacements distributed as $\mathcal{N}(\mathbf{0}, \nu\mathbb{I})$ and covariance $\widetilde{V} = \mathbb{I}$. $\quad\square$

It is interesting to note that despite the construction of the Scrooge distribution in spin systems not *a priori* straightforwardly generalizing to CV systems, the form of the Scrooge distribution Eq. (99) and Gaussian Scrooge distribution Eq. (102) nevertheless bear similarities: both being '$\hat{\rho}$-distortions' of some underlying 'uniform' distribution of states. For the former, it is the uniform, Haar distribution on the Hilbert space, which is normalizable, while in the latter, it is the uniform collection of all unsqueezed coherent states (which does form a resolution of the identity but is not normalizable). We stress though that it was not clear prior to the statement of Theorem 3 that this should have been the case, as a uniform collection of *any* squeezed coherent states also forms a resolution of the identity; one could have also easily argued for a $\hat{\rho}$-distortion of such an ensemble as the limiting form of the PE, though this would give the wrong result.

### D.  Principle of maximum entropy in terms of the mixing entropy

In this section, we consider a possible alternative formulation of the maximum entropy principle for Gaussian CV systems which does yield the Gaussian Scrooge distribution as the limiting form of the projected ensemble.

From the point of view of classical encoding, the Gaussian Scrooge distribution also has the property of maximizing the mixing entropy – a quantity related to the number of classical bits needed to encode the quantum state ensemble. Concretely, the mixing entropy (or the differential entropy) for a multivariate normal distribution is

$$h(p_\Sigma) = \frac{1}{2}\log\det\Sigma + C, \tag{115}$$

where $C$ is a constant depending on the choice of Lebesgue measure. Since the Gaussian projected ensemble discussed in the main text has the form $\mathcal{E}_{\Sigma,\widetilde{V}} = \{p_\Sigma(\mathbf{r})d\mathbf{r}, |\psi(\mathbf{r},\widetilde{V})\rangle\}$, where $\widetilde{V}$ is the common covariance matrix for the ensemble, the only random information needed to encode classically is the random displacement $\mathbf{r}$ for each sample drawn from the ensemble. Hence, asymptotically the number of classical bits needed to encode the Gaussian projected ensemble is captured by the mixing entropy of $p_\Sigma$.

If we consider maximizing the mixing entropy subject to the constraint that $2\Sigma + \widetilde{V} = (1 + 2\nu)\mathbb{I}$, and that $\widetilde{V}$ represents a pure Gaussian state, we would again obtain

$$\arg\max_\Sigma h(p_\Sigma) = \arg\max_\Sigma \det\Sigma = \nu\mathbb{I}. \tag{116}$$

Then $\widetilde{V} = \mathbb{I}$. This is the Gaussian Scrooge distribution.

However, we argue in favor of using the ensemble entropy as the appropriate information entropy to maximize, as opposed to the mixing entropy, even though it does work in the present case. For example, it is known that maximizing the mixing entropy in a spin system (equivalently, the relative entropy of a pure state distribution to the uniform one $-D_{\mathrm{KL}}(p(\psi)\|\mathrm{Haar})$) with the constraint of energy conservation does not produce the expected finite-temperature Scrooge ensemble seen in numerics, while the use of the ensemble entropy does. We attribute this to the fact the mixing entropy is sensitive only to the information contained in the distribution $p(\psi)$ of the projected ensemble and neglects the information contained within the quantum states themselves due to measurements, while the ensemble entropy captures both.

### E.  Complex structure induced by Gaussian state and Williamson decomposition

For completeness, we include a discussion on the complex structure of positive matrices that is necessary to understand Theorem 2 of Holevo [21] (Lemma 5 above). An operator $J$ on the $2n$-dimensional phase space is said to be an operator of complex structure if

$$J^2 = -I_{2n}, \tag{117}$$

where $\Omega$ is evidently an exemplary case of such operators, since $\Omega^{-1} = -\Omega = \Omega^T$. As we will explicitly show below, every Gaussian state with covariance matrix $V$ can induce a complex structure on the phase space, denoted by $J_V$.

Denote the symplectic bilinear form as $\omega(\mathbf{x}, \mathbf{y}) \equiv \mathbf{x}^T\Omega\mathbf{y}$, and consider the inner product $V(\mathbf{x}, \mathbf{y}) \equiv \mathbf{x}^T V\mathbf{y}$ induced by the positive-definite real matrix $V$. Since these two bilinear forms are both non-degenerate, there exists a unique invertible operator $A = \Omega^{-1}V$ that satisfies

$$V(\mathbf{x}, \mathbf{y}) = \omega(\mathbf{x}, A\mathbf{y}), \tag{118}$$

which holds for any $\mathbf{x}, \mathbf{y} \in \mathbb{R}^{2n}$. Next, consider the natural complexification $\mathbb{C}^{2n}$ of the $2n$-dimensional phase space, and the inner product becomes $V(\mathbf{x}, \mathbf{y}) = \mathbf{x}^\dagger V\mathbf{y}$. In this complex vector space, it can be shown that $A^{T_V} = \Omega V = -A$, where $A^{T_V}$ denotes the adjoint of $A$ under the $V$-inner product, defined by $V(A\mathbf{x}, \mathbf{y}) = V(\mathbf{x}, A^{T_V}\mathbf{y})$. Due to this anti-hermiticity, one may then conclude that the operator $A$ has eigenvalues of the form $\{\pm i\lambda_j\}_{j=1}^n$ with $\lambda_j > 0$, and that the eigenvectors come in conjugate pairs as

$$\begin{cases} A(\mathbf{e}_j + i\mathbf{f}_j) = -i\lambda_j(\mathbf{e}_j + i\mathbf{f}_j) \\ A(\mathbf{e}_j - i\mathbf{f}_j) = i\lambda_j(\mathbf{e}_j - i\mathbf{f}_j) \end{cases} \quad\Leftrightarrow\quad \begin{cases} A\mathbf{e}_j = \lambda_j\mathbf{f}_j \\ A\mathbf{f}_j = -\lambda_j\mathbf{e}_j \end{cases}. \tag{119}$$

For simplicity, we assume that $A$ is non-degenerate, and one may thus select a $V$-orthogonal basis $\{\mathbf{e}_j, \mathbf{f}_j\}$ such that $V(\mathbf{e}_j, \mathbf{f}_k) = 0$ and $V(\mathbf{e}_j, \mathbf{e}_k) = V(\mathbf{f}_j, \mathbf{f}_k) = \delta_{jk}\lambda_j$. This forms a symplectic basis as $\omega(\mathbf{e}_j, \mathbf{f}_k) = \delta_{jk}$ and

$\omega(\mathbf{e}_j, \mathbf{e}_k) = \omega(\mathbf{f}_j, \mathbf{f}_k) = 0$. Therefore, the basis transformation $S_{2n \times 2n}$ from the canonical basis to $\{\mathbf{e}_j, \mathbf{f}_j\}$ is symplectic, which we denote explicitly as

$$(\mathbf{e}_1, \cdots, \mathbf{e}_n, \mathbf{f}_1, \cdots, \mathbf{f}_n) = (\hat{q}_1, \cdots, \hat{q}_n, \hat{p}_1, \cdots \hat{p}_n)S, \quad \text{with } S^T \Omega S = \Omega. \tag{120}$$

Consider the polar decomposition of $A$ in the new basis – it consists of two commuting parts:

$$A = |A| J_V = J_V |A|, \tag{121}$$

where $|A| = \sqrt{A^{T_V} A} = \sqrt{A A^{T_V}} = \sqrt{-A^2}$ is positive, and $J_V^{T_V} J_V = J_V J_V^{T_V} = I$. Explicitly, we have

$$\begin{cases} |A|\mathbf{e}_j = \lambda_j \mathbf{e}_j \\ |A|\mathbf{f}_j = \lambda_j \mathbf{f}_j \end{cases} \quad \text{and} \quad \begin{cases} J_V \mathbf{e}_j = \mathbf{f}_j \\ J_V \mathbf{f}_j = -\mathbf{e}_j \end{cases}, \tag{122}$$

and we say $J_V$ is the operator of complex structure induced by the positive-definite $V$. Under the canonical basis representation, we can write the operators as

$$|A| \overset{\text{canonical}}{=\!=\!=\!=} S \Lambda S^{-1}, \quad J_V \overset{\text{canonical}}{=\!=\!=\!=} -S \Omega S^{-1}, \tag{123}$$

where $\Lambda = \text{diag}(\lambda_1, \lambda_2, \cdots, \lambda_n, \lambda_1, \lambda_2, \cdots, \lambda_n)$ is a diagonal matrix. Substituting back to Eq. (121), one can show that

$$S^T V S = \Lambda, \tag{124}$$

which is known as the Williamson decomposition of a positive-definite real matrix $V$. This procedure is also known as a symplectic diagonalization of $V$, and $\{\lambda_j\}_{j=1}^n$ are the symplectic eigenvalues of $V$. Notably, the symplectic eigenvalues of $V$ are unique up to reordering, while the symplectic matrix $S$ diagonalizing $V$ is not unique, as seen by the non-uniqueness of the $V$-orthogonal basis $\{\mathbf{e}_j, \mathbf{f}_j\}$.

Recall that the uncertainty relation satisfied by any Gaussian state is $V \geq \pm i\Omega$, which is equivalent to $\Lambda = S^T V S \geq \pm i S^T \Omega S = \pm i\Omega$, and this amounts to the inequalities

$$\lambda_j \geq 1, \quad \forall j, \tag{125}$$

which is just $\Lambda \geq I$. This inequality is also equivalent to

$$V \geq \Omega J_V. \tag{126}$$

By definition, a pure Gaussian state is a state with covariance matrix $V$ saturating the above uncertainty relations, i.e., for pure states, $V = \Omega J_V$ and the symplectic eigenvalues of $V$ are strictly 1. In the gauge-invariant Gaussian states that [20] considered, the complex structures are always $J_V = \Omega^{-1}$. However, for general cases, the complex structure $J_V$ is not so simple and varies with $V$.

## VII.  WIGNER REPRESENTATION OF $m$-TH MOMENT OF THE GAUSSIAN PROJECTED ENSEMBLE AND GAUSSIAN SCROOGE DISTRIBUTION

In this section, though we never use it in the main text, we provide explicit expressions for the Wigner functions of the $m$-th moment of the Gaussian projected ensemble and Gaussian Scrooge distribution for completeness. The $m$-th moment $\hat{\rho}^{(m)}$ of the Gaussian projected ensemble $\mathcal{E}_G$ is defined as

$$\hat{\rho}^{(m)} = \int d\mathbf{r}_B p(\mathbf{r}_B) \hat{\rho}^{\otimes m}(\mathbf{r}_A(\mathbf{r}_B), \widetilde{V}_A) \tag{127}$$

where $\rho(\mathbf{r}_A(\mathbf{r}_B)) = |\psi(\mathbf{r}_A(\mathbf{r}_B), \widetilde{V}_A)\rangle\langle\psi(\mathbf{r}_A(\mathbf{r}_B), \widetilde{V}_A)|$. The Wigner characteristic function of $\hat{\rho}^{(m)}$ is given by

$$\chi^{(m)}(\boldsymbol{\xi}_1, ..., \boldsymbol{\xi}_m) = \int d\mathbf{r}_B p(\mathbf{r}_B) \text{Tr}(\hat{\rho}^{\otimes m}(\mathbf{r}_A(\mathbf{r}_B), \widetilde{V}_A) \hat{D}(\boldsymbol{\xi}_1) \otimes ... \otimes \hat{D}(\boldsymbol{\xi}_m)) \tag{128}$$

$$= \int d\mathbf{r}_B \frac{e^{-\boldsymbol{\xi}_B^T (V_B + \sigma_B)^{-1} \boldsymbol{\xi}_B}}{\pi^{(2n-2k)} \sqrt{\text{Det}(V_B + \sigma_B)}} \chi_{\rho(\mathbf{r}_A(\mathbf{r}_B), \widetilde{V}_A)}(\boldsymbol{\xi}_1) \cdots \chi_{\rho(\mathbf{r}_A(\mathbf{r}_B), \widetilde{V}_A)}(\boldsymbol{\xi}_m), \tag{129}$$

where the characteristic function of the single copy density matrix $\hat{\rho}(\mathbf{r}_B)$ is given by

$$\chi_{\rho(\mathbf{r}_A(\mathbf{r}_B),\tilde{V}_A)}(\boldsymbol{\xi}_i) = e^{-\frac{1}{4}\boldsymbol{\xi}_i^T \Omega^T \tilde{V}_A \Omega \boldsymbol{\xi}_i + i\Omega \mathbf{r}_A(\mathbf{r}_B)^T \boldsymbol{\xi}_i}. \tag{130}$$

Integrating out $\boldsymbol{\xi}_B$, we obtain

$$\chi^{(m)}(\boldsymbol{\xi}_1,...,\boldsymbol{\xi}_m) = e^{-\sum_i \frac{1}{4}\boldsymbol{\xi}_i^T \Omega^T \tilde{V}_A \Omega \boldsymbol{\xi}_i} \int d\mathbf{r}_B \frac{e^{-\mathbf{r}_B^T(V_B+\sigma_B)^{-1}\mathbf{r}_B}}{\pi^{2n-2k}\sqrt{\text{Det}(V_B+\sigma_B)}} e^{i(\Omega \mathbf{r}_A(\mathbf{r}_B))^T \sum_i \boldsymbol{\xi}_i} \tag{131}$$

$$= e^{-\sum_i \frac{1}{4}\boldsymbol{\xi}_i^T \Omega^T \tilde{V}_A \Omega \boldsymbol{\xi}_i - \frac{1}{4}(\sum_i \boldsymbol{\xi}_i)^T \Omega^T V_{AB}(V_B+\sigma_B)^{-1}V_{AB}^T \Omega(\sum_j \boldsymbol{\xi}_j)} \tag{132}$$

$$= e^{-\frac{1}{4}\boldsymbol{\Xi}^{(m)T}(\Omega^{\oplus m})^T V^{(m)} \Omega^{\oplus m} \boldsymbol{\Xi}^{(m)}}, \tag{133}$$

where

$$\boldsymbol{\Xi}^{(m)} = (\boldsymbol{\xi}_1^T, \boldsymbol{\xi}_2^T, ..., \boldsymbol{\xi}_m^T)^T, \tag{134}$$

$$V^{(m)} = \begin{pmatrix} V_A & ... & V_{AB}(V_B+\sigma_B)^{-1}V_{AB}^T \\ \vdots & \ddots & \vdots \\ V_{AB}(V_B+\sigma_B)^{-1}V_{AB}^T & ... & V_A \end{pmatrix}. \tag{135}$$

Notice that the Wigner characteristic $\hat{\rho}^{(m)}$ is again a Gaussian function, which implies its Wigner function is too. We thus conclude that the higher moments $\hat{\rho}^{(m)}$ of the Gaussian projected ensemble is a mixed Gaussian state which has a zero first moment and covariance matrix $V^{(m)}$. By similar calculation, the Gaussian Scrooge distribution has zero first moment and covariance matrix

$$V_{\text{GSD}}^{(m)} = \begin{pmatrix} (2\nu+1)\mathbb{I}_A & ... & 2\nu\mathbb{I}_A \\ \vdots & \ddots & \vdots \\ 2\nu\mathbb{I}_A & ... & (2\nu+1)\mathbb{I}_A \end{pmatrix}. \tag{136}$$



\end{document}